%% file: paper.tex
\documentclass[sigconf]{acmart}

\usepackage{booktabs} 
\usepackage{hyperref}
\usepackage{listings}
\usepackage[inline]{enumitem}
\usepackage{tabularx}
\usepackage{booktabs}			 	
\usepackage{cleveref}				  
\usepackage{graphicx}				 
\usepackage{multirow}				 
\usepackage[export]{adjustbox} 	
\usepackage{xspace}
\usepackage{wasysym}				
\usepackage{xparse}					  
\usepackage{threeparttable}
\usepackage{subcaption}
\usepackage{calc}						
\usepackage[normalem]{ulem}

\NewDocumentCommand{\rot}{O{90} O{1em} m}{\makebox[#2][l]{\rotatebox{#1}{#3}}}%

\include{preamble}

\usepackage{balance}

\def\BibTeX{{\rm B\kern-.05em{\sc i\kern-.025em b}\kern-.08emT\kern-.1667em\lower.7ex\hbox{E}\kern-.125emX}}

\setcopyright{acmlicensed}

\begin{document}
\fancyhead{}
	
\title{AdVersarial: Perceptual Ad Blocking meets\\ Adversarial Machine Learning} 

%
\author{Florian Tram\`er}
\email{tramer@cs.stanford.edu}
\affiliation{Stanford University}
\author{Pascal Dupr\'{e}}
\email{s9padupr@stud.uni-saarland.de}
\affiliation{CISPA Helmholtz Center for Information Security}
\author{Gili Rusak} 
\email{gili@stanford.edu}
\affiliation{Stanford University}
\author{Giancarlo Pellegrino}
\email{gpellegrino@cispa.saarland}
\affiliation{Stanford University, CISPA Helmholtz Center for Information Security}
\author{Dan Boneh}
\email{dabo@cs.stanford.edu}
\affiliation{Stanford University}

\copyrightyear{2019} 
\acmYear{2019} 
\acmConference[CCS '19]{2019 ACM SIGSAC Conference on Computer and Communications Security}{November 11--15, 2019}{London, United Kingdom}
\acmBooktitle{2019 ACM SIGSAC Conference on Computer and Communications Security (CCS '19), November 11--15, 2019, London, United Kingdom}
\acmPrice{15.00}
\acmDOI{10.1145/3319535.3354222}
\acmISBN{978-1-4503-6747-9/19/11}

\begin{abstract}
	\emph{Perceptual ad-blocking} is a novel approach that detects 
	online advertisements based on their visual content. 
	Compared to traditional filter lists, the use of perceptual signals is believed 
	to be less prone to an arms race with web publishers and ad networks.
	We demonstrate that this may not be the case. We 
	describe attacks on multiple perceptual ad-blocking techniques, and 
	unveil a new arms race that likely disfavors ad-blockers.  
	Unexpectedly, perceptual ad-blocking can also introduce
	new vulnerabilities that let an attacker bypass web 
	security boundaries and mount DDoS attacks.
	
	We first analyze the design space of perceptual ad-blockers and present a unified architecture that incorporates prior academic and commercial work.
	We then explore a variety of attacks on the ad-blocker's detection pipeline, that enable publishers or ad networks to evade or detect ad-blocking, and at times even abuse its high privilege level to bypass web security boundaries.
	
	On one hand, we show that perceptual ad-blocking must visually classify rendered web content to escape an arms race centered on obfuscation of page markup. On the other, we present a concrete set of attacks on visual ad-blockers by constructing \emph{adversarial examples} in a real web page context. For seven ad-detectors, we create perturbed ads, ad-disclosure logos, and native web content that misleads perceptual ad-blocking with 100\% success rates. 
	In one of our attacks, we demonstrate how a malicious user can upload adversarial content, such as a perturbed image in a Facebook post, that fools the ad-blocker into removing another users' non-ad content.
	
	Moving beyond the Web and visual domain, we also build adversarial examples for AdblockRadio, an open source radio client that uses machine learning to detects ads in raw audio streams.
\end{abstract}



\maketitle

\input{introduction}
\input{background}

\input{taxonomy}

\input{attacks_new}
\input{discussion}
\input{relwork}

\input{conclusion}

\bibliographystyle{ACM-Reference-Format}
\bibliography{biblio}

\appendix 
\input{appendix_detection}

\input{appendix_yolo}
\input{appendix_extra}

\end{document}

%% file: preamble.tex
\usepackage{xcolor}

\newenvironment{proof-sketch}{\noindent{\bf Sketch of Proof}\hspace*{1em}}{\qed\bigskip}
\newenvironment{proof-idea}{\noindent{\bf Proof Idea}\hspace*{1em}}{\qed\bigskip}
\newenvironment{proof-of-lemma}[1]{\noindent{\bf Proof of Lemma #1}\hspace*{1em}}{\qed\bigskip}
\newenvironment{proof-attempt}{\noindent{\bf Proof Attempt}\hspace*{1em}}{\qed\bigskip}

{\makeatletter
 \gdef\xxxmark{%
   \expandafter\ifx\csname @mpargs\endcsname\relax 
     \expandafter\ifx\csname @captype\endcsname\relax 
       \marginpar{\textcolor{red}{xxx~}}
     \else
       \textcolor{red}{xxx~}
     \fi
   \else
     \textcolor{red}{xxx~}
   \fi}
 \gdef\xxx{\@ifnextchar[\xxx@lab\xxx@nolab}
 \long\gdef\xxx@lab[#1]#2{{\bf [\xxxmark \textcolor{red}{#2} ---{\sc #1}]}}
 \long\gdef\xxx@nolab#1{{\bf [\xxxmark \textcolor{red}{#1}]}}
}

\newcommand{\R}{\mathbb{R}}

\newcommand{\calL}{\ensuremath{\mathcal{L}}}

\newenvironment{myitemize}
{\begin{itemize}[leftmargin=5mm,itemsep=1mm,topsep=2pt,after=\vspace{0.5\baselineskip}]}
	{\end{itemize}}

\newenvironment{myenumerate}
{\begin{enumerate}[leftmargin=5mm,itemsep=2mm,topsep=2pt,after=\vspace{\baselineskip}]}
	{\end{enumerate}}

\newcommand{\iframe}{\texttt{iframe}\xspace}
\newcommand{\img}{\texttt{img}\xspace}

\definecolor{editorGray}{rgb}{0.95, 0.95, 0.95}
\definecolor{editorOcher}{rgb}{0.25, 0.25, 1}
\definecolor{editorGreen}{rgb}{0, 0.3, 0} 
\lstdefinestyle{basic}{%
	backgroundcolor=\color{editorGray},
	basicstyle={\scriptsize\ttfamily},   
	frame=tb,
	xleftmargin=.01\columnwidth, 
	xrightmargin=.01\columnwidth,
	framesep=0pt,
	framextopmargin=2pt,
	stringstyle=\color{editorOcher},
	language=html,
	alsodigit={.:;},
	tabsize=2,
	showtabs=false,
	showspaces=false,
	showstringspaces=false,
	extendedchars=true,
	breaklines=true,        
}

\lstdefinelanguage{HTMLSpecial}{
	morekeywords={span, class, a},
	morekeywords=[2]{Sp,on,so,red},
	keywordstyle=[2]{\color{red}\bfseries},
	morestring=[s]{"}{"},
}

\lstdefinelanguage{CSS}{
	morekeywords={background-image,width,height,top,left,
		position,z-index,opacity,pointer-events,url,fixed,
		none,font-size},
	morekeywords=[2]{0,100\%,10000,0.01,.},
	keywordstyle=[2]{\color{purple}},
	morekeywords=[3]{\#overlay,.c2},
	keywordstyle=[3]{\color{editorGreen}\bfseries},
	morestring=[s]{"}{"},
	sensitive,
	alsoletter={-,\#,0,1,2,3,4,5,6,7,8,9,\%,.},
}

%% file: introduction.tex
\section{Introduction}

Online advertising is a contentious facet of the Web. Online ads generate over \$200 billion in value~\cite{PWCoutlook}, but many Internet users perceive them as intrusive or malicious~\cite{Pujol:2015:AUA:2815675.2815705,kontaxis2015tracking,Xing:2015:UMT:2736277.2741630, li2012knowing}. The growing use of ad-blockers such as Adblock Plus and uBlock~\cite{adblockplus,ublock} has sparked a fierce arms race with publishers and advertising networks. Current ad-blockers maintain large crowdsourced lists of ad \emph{metadata}---such as page markup and URLs. In turn, publishers and ad networks (including Facebook~\cite{ublock-forum,fbadblockplus} and 30\% of the Alexa top-10K list~\cite{zhu2018measuring}) continuously adapt and deploy small changes to web page code in an effort to evade, or detect ad-blocking.

\paragraph{Towards visual ad-blocking}
This arms race has prompted ad-blockers to search for more robust signals within ads' visual \emph{content}, as altering these would affect user experience. One such signal is the presence of ad-disclosures such as a ``Sponsored'' caption or the AdChoices logo~\cite{DAAprinciples}), which many ad-networks add for transparency~\cite{DAAprinciples}. Storey et al.~\cite{storey} proposed Ad-Highlighter~\cite{adhighlighter}, the first \emph{perceptual ad-blocker} that detects ad-disclosures by combining web-filtering rules and computer vision techniques. Motivated by the alleged superior robustness of perceptual techniques~\cite{storey}, popular ad-blockers now incorporate similar ideas. For example, Adblock Plus supports image-matching filters~\cite{adblockplus}, while uBlock crawls Facebook posts in search for ``Sponsored'' captions~\cite{ublock}. 

However, as proposed perceptual ad-blockers still partially use markup as a proxy for ads' visual content, they appear insufficient to end the ad-blocking arms race. 
For example, Facebook routinely evades uBlock Origin using increasingly complex \emph{HTML obfuscation} for the ``Sponsored'' captions (see~\cite{ublock-forum}). Ad-Highlighter's computer vision pipeline is also vulnerable to markup tricks such as image fragmentation or spriting (see Figure~\ref{fig:adchoices-sprites}). Escaping the arms race over markup obfuscation requires perceptual ad-blockers to move towards operating on \emph{rendered} web content. This is exemplified in Adblock Plus' Sentinel project~\cite{sentinel}, that uses deep learning to detect ads directly in web page screenshots. On a similar note, Percival~\cite{percival} is a recently proposed ad-blocker that adds a deep learning ad-classifier into the rendering pipeline of the Chromium and Brave browsers. While these approaches might bring an end to the current markup-level arms race, our paper shows that visual ad-blocking merely replaces this arms race with a new one, involving powerful attacks that directly target the ad-blockers' visual classifier.

\paragraph{A security analysis of perceptual ad-blocking.} 
In this paper, we present the first comprehensive security analysis of perceptual ad-blockers, and challenge the belief that perceptual signals will end the ad-blocking arms race.
To provide a principled analysis of the design space of these nascent ad-blocking techniques, we first propose a general architecture that incorporates and extends existing approaches, e.g., Ad-Highlighter~\cite{adhighlighter,storey}, Sentinel~\cite{sentinel,sentinel-medium} and Percival~\cite{percival}. We view perceptual ad-blocking as a classification pipeline, where segmented web data is fed into one of a variety of possible ad (or ad-disclosure) detectors. 

Given this unified view of the design space, we identify and analyze a variety of attacks that affect each step of the ad-classification pipeline. Multiple adversaries---publishers, ad networks, advertisers or content creators---can exploit these vulnerabilities to evade, detect and abuse ad-blockers. Our attacks combine techniques from Web security and from adversarial machine learning~\cite{papernot2016towards}. In particular, we leverage visual \emph{adversarial examples}~\cite{szegedy2013intriguing}, slightly perturbed images that fool state-of-the-art classifiers.

\paragraph{Web attacks on perceptual ad-blockers.}
First, we show that ad-blocking approaches that combine traditional markup filters and visual signals remain vulnerable to the same attacks as current filter-lists. HTML obfuscation of ad-disclosures is already observed today~\cite{ublock-forum}, and we demonstrate similar attacks against Ad-Highlighter's image-matching pipeline (e.g., by fragmenting images). Thus, unless ad-blockers move towards relying on rendered web content (as in Sentinel~\cite{sentinel}), perceptual signals will not end the ongoing markup arms race with ad-networks and publishers.

In addition to visual signals, Storey et al.~\cite{storey} suggest to detect ad-disclosures using \emph{behavioral signals} such as the presence of a link to an ad-policy page. We demonstrate that such signals can lead to serious vulnerabilities (e.g., CSRF, DDoS or click-fraud). Specifically, we show how a Facebook user can trick Ad-Highlighter into making arbitrary web requests in other ad-block users' browsers.

\paragraph{Adversarial examples for ad-classifiers.}
Ad-blockers can counter the above attacks by operating on \emph{rendered} web content. 
The main threat to visual ad-blockers are then adversarial examples, which challenge the core assumption that ML can emulate humans' visual ad-detection. To our knowledge, our attacks are the first application of adversarial examples to a real-world web-security problem. \footnote{Gilmer et al.~\cite{gilmer2018motivating} argue that the threat model of adversarial examples---in particular the fact that the adversary is restricted to imperceptible perturbations of a given input---is often unrepresentative of real security threats. Perceptual ad-blocking is a perfect example where this threat model is entirely appropriate. The ad-blocker's adversaries---who have white-box access to its classifier---want to evade it on specific inputs (e.g., an ad-network cannot ``sample'' new ads until it finds one that evades the ad-blocker), with attacks that the user should be oblivious to.}

We rigorously assess the threat of adversarial examples on \emph{seven} visual ad-classifiers: 
Two computer-vision algorithms (perceptual hashing and OCR) used in Ad-Highlighter~\cite{storey}; the ad-classification neural networks used by Percival~\cite{percival} and~\cite{hussain2017automatic}; a canonical feature matching model based on the \emph{Scale-Invariant Feature Transform} (SIFT)~\cite{lowe2004distinctive}; and two object detector networks emulating Sentinel~\cite{sentinel}. 
For each model, we create imperceptibly perturbed ads, ad-disclosure or native content, that either evade the model's detection or falsely trigger it (as a means of detecting ad-blocking). 


Among our contributions is a new evasion attack~\cite{ilyas2018black, salimans2017evolution} on SIFT~\cite{lowe2004distinctive} that is conceptually simpler than prior work~\cite{hsu2009secure}.

Attacking perceptual ad-blockers such as Sentinel~\cite{sentinel} presents the most interesting challenges. For these, the classifier's input is a screenshot of a web page with contents controlled by different entities (e.g, publishers and ad networks). Adversarial perturbations must thus be encoded into HTML elements that the adversary controls, be robust to content changes from other parties, and scale to thousands of pages and ads. We tackle the adversary's uncertainty about other parties' page contents by adapting techniques used for creating physical adversarial examples~\cite{sharif2016accessorize,athalye2018synthesizing}. We also propose a novel application of \emph{universal adversarial examples}~\cite{moosavi2017universal} to create a \emph{single} perturbation that can be applied at scale to all combinations of websites and ads with near $100\%$ success probability.  

\begin{figure}[t]
	\centering
	{%
		\setlength{\fboxsep}{0pt}%
		\setlength{\fboxrule}{1pt}%
		\fbox{%
			\includegraphics[width=0.9\columnwidth-2pt]{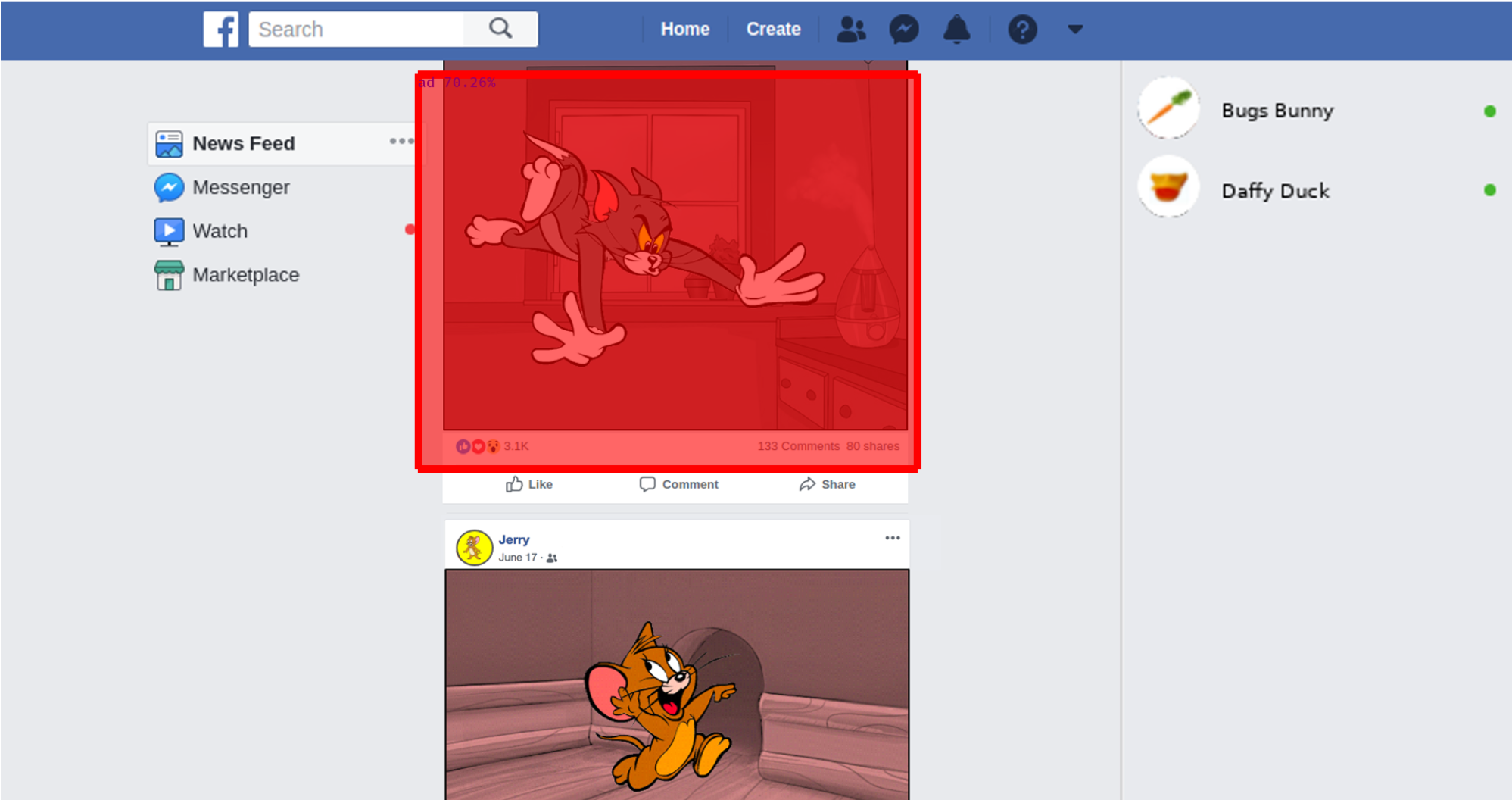}%
		}%
	}%
	\vspace{-0.65em}
	\caption{\textbf{Ad-Blocker Privilege Hijacking.} A malicious user, Jerry, posts adversarial content to Facebook that fools a perceptual ad-blocker similar to Sentinel~\cite{sentinel} into marking Tom's benign content as an ad (red box) and blocking it in every user's browser.\\[-2em]}
	\label{fig:tom-jerry}
\end{figure}

We further show that adversarial examples enable new attacks, wherein malicious content from one user can hijack the ad-blocker's high privilege to incorrectly block another user's content.  An example is shown in \Cref{fig:tom-jerry}. Here Jerry, the adversary, uploads a perturbed image to Facebook.  That image is placed next to Tom's post, and confuses the ad-blocker into classifying Tom's benign post as an ad, and incorrectly blocking it.  Hence, a malicious post by one user can cause another user's post to get blocked.

Moving beyond the Web and visual domain, we build imperceptible audio adversarial examples for AdblockRadio~\cite{adblockradio}, a radio ad-blocker that uses ML to detect ads in raw audio streams.

\paragraph{Outlook.} While visual ad-classification of rendered web content is both sufficient and necessary to bring an end to the arms race around page markup obfuscation, we show that this merely replaces one arms race with a new one centered on adversarial examples. Our attacks are not just a first step in this new arms race, where ad-blockers can easily regain the upper hand. Instead, they describe an inherent difficulty with the perceptual ad-blocking approach, as ad-blockers operate in essentially the \emph{worst threat model for visual classifiers}.
Their adversaries prepare (offline) {digital} attacks to evade or falsely trigger a known \emph{white-box} visual classifier running inside the ad-blocker. In contrast, the ad-blocker must resist these attacks while operating under strict real-time constraints.

Our study's goal is not to downplay the merits of ad-blocking, nor discredit the perceptual ad-blocking philosophy. Indeed, ML might one day achieve human-level perception. Instead, we highlight and raise awareness of the inherent vulnerabilities that arise from instantiating perceptual ad-blockers with existing ML techniques. 

\vspace{0.1cm}
\noindent\mbox{\textbf{Contributions.} This paper makes the following contributions:}

\begin{myitemize}
    \item We conduct a detailed security analysis of perceptual ad-blocking;


    \item We present nine general classes of attacks against the various components of the perceptual ad-blocking pipeline;

    \item We evaluate adversarial examples for eight ad classifiers (seven visual, one audio). We make novel use of transformation-robust~\cite{athalye2018synthesizing} and universal adversarial examples~\cite{moosavi2017universal} to create scalable attacks robust to arbitrary changes in web content.

    \item We release all our data and classifiers, including a new neural network that locates ads in web page screenshots, that may prove useful \emph{in non-adversarial settings}: \url{https://github.com/ftramer/ad-versarial}
\end{myitemize}

%% file: background.tex
\section{Preliminaries and Background}
\label{sec:background}


\subsection{The Online Advertising Ecosystem} 
\label{sec:online_ad}

Online advertising comprises four actors: users, publishers, ad networks, and advertisers. Users browse websites owned or curated by a publisher. Publishers assigns parts of the site's layout to advertisements.  Control of these spaces is often outsourced to an ad network that populates them with advertisers' contents. 

\begin{figure}[!t]
	\centering
	\setlength{\tabcolsep}{2pt}
	\bgroup
	\def\arraystretch{1.25}
	\begin{tabularx}{\columnwidth}{@{}c c c c@{}}
		\multirow{2}{20px}{}
		&
		\multirow{2}{36px}{
			{%
				\setlength{\fboxsep}{2px}%
				\setlength{\fboxrule}{0.1pt}%
				\fbox{%
					\includegraphics[height=23px,valign=M]{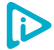}%
				}%
			}%
		}
		&
		\multirow{2}{75px}{
			{%
				\setlength{\fboxsep}{2px}%
				\setlength{\fboxrule}{0.1pt}%
				\fbox{%
					\includegraphics[height=12px,valign=M]{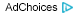}%
				}%
			}%
		}
		&
		{%
			\includegraphics[height=12px,valign=M]{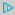}%
		}%
		$\ $
		{%
			\setlength{\fboxsep}{1px}%
			\setlength{\fboxrule}{0.1pt}%
			\fbox{%
				\includegraphics[height=10px,valign=M]{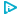}%
			}%
		}%
		$\ $
		{%
			\setlength{\fboxsep}{1.5px}%
			\setlength{\fboxrule}{0.1pt}%
			\fbox{%
				\includegraphics[height=9px,valign=M]{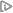}%
			}%
		}%
		\\
		&&&
		{%
			\setlength{\fboxsep}{1px}%
			\setlength{\fboxrule}{0.1pt}%
			\fbox{%
			\includegraphics[height=13px,valign=M]{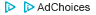}%
			}%
		}%
		\\
		& (a) & (b) & (c)
	\end{tabularx}
	\egroup
	\setlength{\tabcolsep}{6pt}
	\vspace{-1em}
	\caption{\textbf{The AdChoices Logo.} AdChoices is a standard for disclosure of behavioral advertising~\cite{DAAprinciples}. Ads are marked by the icon (a), with optional text (b). Despite creative guidelines~\cite{DAA2013guidelines}, many variants of the logo are in use (c).\\[-2em]}
	\label{fig:adchoices-logo}	
\end{figure}

To protect users from deceptive ads, the Federal Trade Commission and similar non-US agencies require ads to be clearly recognizable~\cite{edelman2009false}. These provisions have also spawned industry self-regulation, such as the \emph{AdChoices} standard~\cite{DAAprinciples} (see Figure~\ref{fig:adchoices-logo}).

\subsection{Perceptual Ad-Blocking} 

Perceptual ad-blocking aims at identifying ads from their content, rather than from ad \emph{metadata} such as URLs and markup.
The insight of Storey et al.~\cite{storey} is that many ads are explicitly marked---e.g., via a ``Sponsored'' link or the AdChoices logo---to comply with regulations on deceptive advertising.
They developed Ad-Highlighter~\cite{adhighlighter}, an ad-blocker that detects ad-disclosures using different perceptual techniques: (i) textual searches for ``Sponsored'' tags, (ii) fuzzy image search and OCR to detect the AdChoices logo, and (iii) ``behavioral'' detection of ad-disclosures by identifying links to ad-policy pages. 
Ad-blockers that rely on perceptual signals are presumed to be less prone to an arms race, as altering these signals would affect user experience or violate ad-disclosure regulations~\cite{storey}.

Perceptual ad-blocking has drawn the attention of major ad-blockers, that have integrated visual signals into their pipelines. For example, uBlock blocks Facebook ads by detecting the ``Sponsored'' caption. Adblock Plus has added support for image-matching rules, which are easily extended to fuzzy image search~\cite{adblock-snippet}.

The above perceptual ad-blocking approaches still rely on some markup data as a proxy for ads' visual content. This has prompted an ongoing arms race between Facebook and uBlock (see~\cite{ublock-forum}) where the former continuously obfuscates the HTML tags that render its ``Sponsored'' tag---a process that is invisible to the user.
This weakness is fundamental to perceptual approaches that rely on signals with an indirect correspondence to ads' rendered content. This insight led Adblock Plus to announce the ambitious goal of detecting ads directly from rendered web pages---with no reliance on markup---by leveraging advances in image classification.
Their Sentinel~\cite{sentinel} project uses an object-detection neural network to locate ads in raw Facebook screenshots. The recently released Percival project~\cite{percival} targets a similar goal, by embedding a deep-learning based ad-blocker directly into Chromium's rendering engine.

\subsubsection{Design and Goals.} 
Ad-blockers are client-side programs running in browsers at a high privilege level. They can be implemented as browser extensions or integrated in the browser. We ignore DNS ad-blockers (e.g., Pi-hole) as these cannot use perceptual signals. \footnote{While our work focuses on the desktop browser setting, perceptual ad-blocking might also prove useful in the mobile domain. Current mobile ad-blockers are often part of a custom browser, or act as web proxies---an insufficient approach for native apps that prevent proxying using certificate pinning. Instead, a perceptual ad-blocker (potentially with root access) could detects ads directly from app screenshots 
}

The goal of ad-blockers is to identify and hide ads, while guarding against website breakage~\cite{breakage} resulting from the removal of functional content. 
As opposed to network-level filters, perceptual signals only apply to downloaded Web content and are thus unsuitable for some secondary goals of ad-blockers, such as bandwidth saving or blocking of user tracking and malvertising~\cite{kontaxis2015tracking, Pujol:2015:AUA:2815675.2815705, li2012knowing, Xing:2015:UMT:2736277.2741630}. 

Ad-blockers may strive to remove ads without being detected by the publisher. For example, many websites try to detect ad-blockers~\cite{mughees2016first} and take according action (e.g., by asking users to disable ad-blockers).  
As perceptual ad-blockers do not interfere with web requests, they are undetectable by the web-server~\cite{storey}. However, the publisher's JavaScript code can try to detect ad-blockers by observing changes in the DOM page when hiding ads. 

Finally, perceptual ad-blockers have strict timing constraints, and should process a web page and detect ads in close to real-time. 

\subsubsection{Algorithms for Visual Ad Classification}

The identification of ads or ad-disclosures can be achieved using a variety of computer vision techniques. Below, we describe existing approaches.





\begin{myitemize}
	\item \emph{Template matching.} Ad-Highlighter detects the AdChoices logo by comparing each image in a page to a template using \emph{average hashing}: for each image, a hash is produced by resizing the image to a fixed size and setting the $i$\textsuperscript{th} bit in the hash to $1$ if the $i$\textsuperscript{th} pixel is above the mean pixel value. An image matches the template if their hashes have a small Hamming distance.
	
	A more robust template matching algorithm is the Scale Invariant Feature Transform (SIFT)~\cite{lowe2004distinctive}, which creates an image hash from detected ``keypoints'' (e.g., edges and corners).
	
	\item \emph{Optical Character Recognition.}
	To detect the rendered text inside the AdChoices logo, Ad-Highlighter uses the open-source Tesseract OCR system~\cite{tesseract}. Tesseract splits an image into overlapping frames and transcribes this sequence using a neural network. Ad-Highlighter matches images for which the OCR output has an edit-distance with ``AdChoices'' below $5$.
	
	\item \emph{Image Classification.} Albeit not in an ad-blocking context, Hussain et al.~\cite{hussain2017automatic} have demonstrated that neural networks could be trained to distinguish images of ads from non-ads (without the presence of any explicit ad-disclosures). The Percival project trained a similar neural network to classify image frames in real-time within Chromium's rendering pipeline~\cite{percival}.
	
	\item \emph{Object Detection.}  Sentinel~\cite{sentinel} detects ads in rendered web pages using an object detector network based on YOLOv3~\cite{yolo3}. The network's output encodes locations of ads in an image. The YOLOv3~\cite{yolo3} model outputs bounding box coordinates and confidence scores for $B=10{,}647$ object predictions, and retains those with confidence above a threshold $\tau= 0.5$. 
\end{myitemize}

\subsection{Threat Model and Adversaries}
\label{ssec:adversaries}

We adopt the terminology of adversarial ML~\cite{papernot2016towards}, where the defenders are users of a classifier (the ad-blocker) that its adversaries (e.g., ad networks or publishers) are trying to subvert.

Publishers, ad networks, and advertisers have financial incentives to evade or detect ad-blockers. We assume that publishers and ad networks are rational attackers that abide by regulations on online advertising, and also have incentives to avoid actively harming users or disrupting their browsing experience.
As shown in prior work~\cite{Xing:2015:UMT:2736277.2741630,pellegrino2015cashing}, this assumption fails to hold for advertisers, as some have abused ad-networks for distributing malware. We assume that advertisers and content creators (e.g., a Facebook user) may try to actively attack ad-block users or other parties.

As ad-blockers are client-side software, adversaries can download and inspect their code offline. However, we assume that adversaries do not know \emph{a priori} whether a user is running an ad-blocker.

\paragraph{Attacking ad-blockers.} The primary adversarial goal of publishers, ad-networks and advertisers is to evade the ad-blocker's detection and display ads to users. These adversaries may modify the structure and content of web pages or ads to fool the ad-detector. 

Alternatively, the ad-blocker's adversaries may try to detect its presence, to display warnings or deny access to the user. A common strategy (used by $30\%$ of publishers in the Alexa top-10k) adds fake ad-content (honeypots) to a page and uses JavaScript to check if the ads were blocked~\cite{zhu2018measuring}. This practice leads to an orthogonal arms race on ad-block detection~\cite{mughees2016first, mughees2017detecting, nithyanand2016adblocking} (see Appendix~\ref{apx:detection}).

Adversaries may also try to abuse ad-blockers' behaviors to degrade their usability (e.g., by intentionally causing site-breakage or slow performance). The viability of such attacks depends on the adversary's incentives to avoid disrupting ad-block users' browsing experience (e.g., Facebook adds honeypots to regular user posts to cause site-breakage for ad-block users~\cite{ublock-forum}). 

Finally, attackers with no ties to the online advertisement ecosystem may try to hijack an ad-blocker's high privilege-level in other users' machines. Such attackers can act as advertisers or content creators to upload malicious content that exploits an ad-blocker's vulnerabilities. \Cref{fig:tom-jerry} shows one example of such an attack, where a malicious Facebook user uploads content that tricks the ad-blocker into hiding an honest user's posts. We will also show how Facebook users can exploit Ad-Highlighter's behavioral ad-blocking to trigger arbitrary web requests in other users' browsers.

\subsection{Adversarial Examples}
\label{ssec:adv_examples}






\begin{figure*}[!htb]
	\centering
	\includegraphics[width=0.95\textwidth]{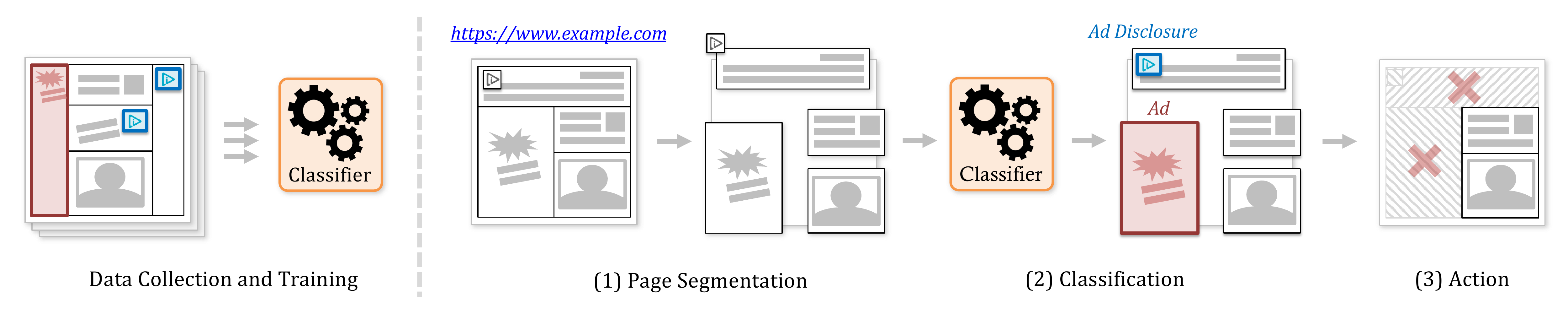}
	\vspace{-1.5em}
	\caption{\textbf{The Architecture of a Perceptual Ad-Blocker.} In the offline phase, an ad-classifier is trained on web data. In the online phase, the ad-blocker segments visited pages (1), classifies individual elements (2), and renders the user's ad-free viewport (3).\\[-2em]}
	\label{fig:architecture}
\end{figure*}

The attacks presented in this paper combines techniques from Web security and from adversarial machine learning. In particular, we leverage adversarial examples~\cite{szegedy2013intriguing} to fool perceptual ad-classifiers.

An adversarial example for an input $x$ of a model $f$ is an input $\hat{x}=x+\delta$, where $\delta$ is a ``small'' perturbation such that $\hat{x}$ is misclassified with high confidence. 
We will consider perturbations with small $\ell_2$ norm (Euclidean) or $\ell_\infty$ norm (maximum per-pixel change). 
To cause a  model $f$ to misclassify $x + \delta$ we would minimize the confidence of $f(x + \delta)$ in the true class, while also keeping $\delta$ small. This is often achieved by minimizing a differentiable \emph{loss function} $\calL(x + \delta)$ that acts as a proxy for the adversarial goal. 

An effective algorithm for finding adversarial examples is \emph{Projected Gradient Descent} (PGD)~\cite{madry2017towards,kurakin2016adversarial}. Given an allowable perturbation set (e.g., $||\delta||_\infty \leq \epsilon$), we repeatedly update $\delta$ in the gradient direction $-\nabla_\delta \calL(x+\delta)$ and project back onto the allowable set.

In some cases, we will want perturbations to be re-usable so that an attack can scale to a large number of websites or ads. A perturbation that can be re-used for many different inputs is called a \emph{universal} adversarial example~\cite{moosavi2017universal}. It is usually created by jointly minimizing $\sum_i\mathcal{L}(x^{(i)} + \delta)$ over many inputs $x^{(i)}$, for a common $\delta$.

%% file: taxonomy.tex

\section{Designing Perceptual Ad-Blockers}
\label{sec:percept_adb}
To analyze the security of perceptual ad-blockers, 
we first propose a unified architecture that incorporates and extends prior and concurrent work (e.g., Ad-Highlighter~\cite{storey}, visual filter-lists~\cite{adblock-snippet}, Sentinel~\cite{sentinel}, and 
the recent Percival patch for Chromium's rendering engine). We explore different ways in which ad-blockers can integrate perceptual signals, and identify a variety of computer vision and ML  techniques that can be used to visually identify ads.

To simplify exposition, we restrict our analysis to ad-blockers that only rely on perceptual signals. In practice, these signals are likely to be combined with existing filter lists (as in uBlock~\cite{ublock-forum} or Adblock Plus~\cite{adblock-snippet}) but the details of such integrations are orthogonal to our work. We note that an ad-blocker that combines perceptual signals with filter lists inherits the vulnerabilities of both, so our security analysis applies to these hybrid approaches as well.

\subsection{General Architecture}
\label{ssec:architecture}

A perceptual ad-blocker is defined by a collection of offline and online steps, with the goal of creating, maintaining and using a classifier to detect ads. \Cref{fig:architecture} shows our unified architecture for perceptual ad-blockers. The ad-blocker's core visual classifier can range from classical computer vision as in Ad-Highlighter~\cite{adhighlighter} to large ML models as in Sentinel~\cite{sentinel}.

The classifier may be trained using labeled web data, the type and amount of which varies by classifier. Due to continuous changes in web markup, ad-blockers may need regular updates, which can range from extending existing rules (e.g., for Ad-Highlighter~\cite{storey, adhighlighter}) to re-training complex ML models such as Sentinel~\cite{sentinel}.

When deployed by a user, the ad-blocker analyzes data from visited pages to detect and block ads in real-time. Ad detection consists of three main steps. (1) The ad-blocker optionally segments the web page into smaller chunks. (2) A classifier labels each chunk as ad or non-ad content. (3) The ad-blocker acts on the underlying web page based on these predictions (e.g., to remove HTML elements labeled as ads). 
For some ad-classifiers, the segmentation step may be skipped. For example, Sentinel~\cite{sentinel} uses an object-detection network that directly processes full web page screenshots.

Ad-Highlighter's use of behavioral signals (i.e., recognizing ad-disclosures by the presence of a link to an ad-policy page) can be seen as a special type of classifier that may interact with segmented Web elements (e.g., by clicking and following a link).

\subsection{Approaches to Ad Detection}
\label{ssec:approaches}

\begin{figure}[t]
	\centering
	\vspace{-1em}
	\includegraphics[width=0.75\columnwidth]{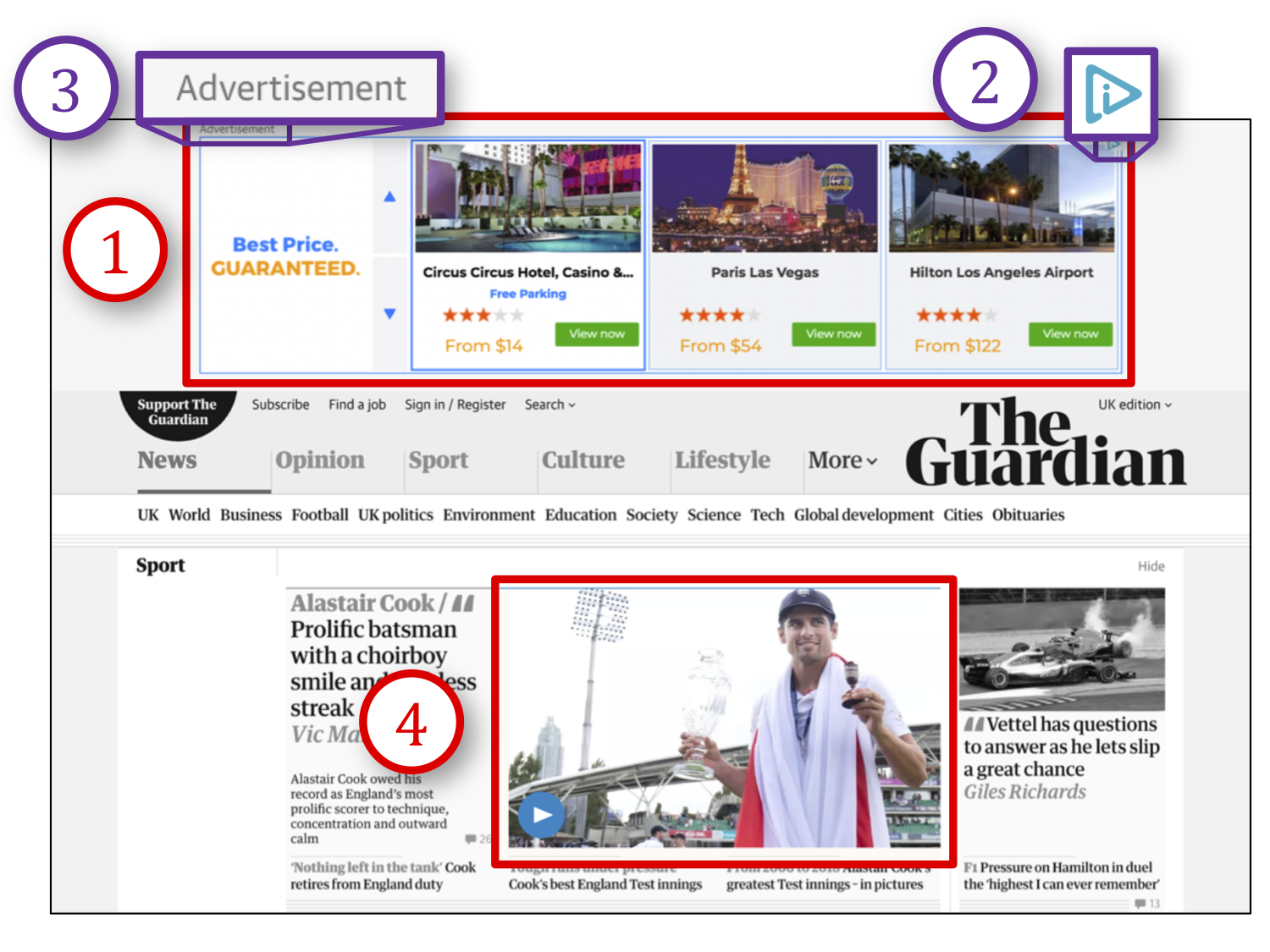}
	\vspace{-1.5em}
	\caption{\textbf{Perceptual Ad-Blocking Elements.} An ad (box $\#1$) is displayed in an \texttt{iframe}, that contains an AdChoices icon (box $\#2$). A custom ad-disclosure from the publisher is outside the \texttt{iframe} (box $\#3$). Publishers can use $\texttt{iframes}$ to display non-ad content such as videos (box $\#4$).\\[-2em]}
	\label{fig:elements}
\end{figure}

When online, a perceptual ad-blocker's first action is the ``Page Segmentation'' step that prepares inputs for the classifier. 
\Cref{fig:elements} illustrates different possible segmentations. A cross-origin \texttt{iframe} (red box 3) displays an ad and an AdChoices icon (purple box 2). An additional textual ad-disclosure is added by the publisher outside the \texttt{iframe} (purple box 1). Publishers may use \texttt{iframes} to display native content such as videos (e.g., red box 4).

We distinguish three main perceptual ad-blocking designs that vary in the granularity of their segmentation step, and in turn in the choice of classifier and actions taken to block ads.
\begin{myitemize}
\item \emph{Element-based perceptual ad-blockers}, such as Ad-Highlighter, search a page's DOM tree for HTML elements that identify ads, e.g., the AdChoices logo or other ad-disclosures. 

\item \emph{Page-based perceptual ad-blockers}, e.g., Sentinel~\cite{sentinel}, ignore the DOM and classify images of rendered web pages. 

\item \emph{Frame-based perceptual ad-blockers}, e.g., Percival~\cite{percival}, classify rendered content but pre-segment pages into smaller frames.
\end{myitemize}


\subsubsection{Element-based Perceptual Ad-blocking.}
These ad-blockers segment pages into HTML elements that are likely to contain ad-disclosures.  The segmentation can be coarse (e.g., Ad-Highlighter extracts all \img tags from a page) or use custom filters as in Adblock Plus' image search~\cite{adblock-snippet} or Ublock's Facebook filters~\cite{ublock-forum}.

For textual ad-disclosures (e.g., Facebook's ``Sponsored'' tag) the classification step involves trivial string matching. Facebook is thus deploying HTML obfuscation that targets an ad-blocker's ability to find these tags~\cite{ublock-forum}. This ongoing arms race calls for the use of visual (markup-less) detection techniques. Ad-disclosure logos (e.g., the AdChoices icon) can be visually classified using \emph{template matching}. Yet, due to many small variations in ad-disclosures in use, \emph{exact} matching (as in Adblock Plus~\cite{adblock-snippet}) is likely insufficient~\cite{storey}.
Instead, Ad-Highlighter uses \emph{perceptual hashing} to match all \img elements against the AdChoices logo. Ad-Highlighter also uses supervised ML---namely Optical Character Recognition (OCR)---to detect the ``AdChoices'' text~\cite{adhighlighter}. Once an ad-disclosure is identified, the associated ad is found using custom rules (e.g., when Ad-Highlighter finds an AdChoices logo, it blocks the parent \texttt{iframe}).

Storey et al.~\cite{storey} further suggest to detect ads through \emph{behavioral signals} that capture the ways in which users can interact with them, e.g., the presence of a link to an ad-policy page.

\subsubsection{Frame-based Perceptual Ad-blocking.} 
The above element-based approaches require mapping elements in the DOM to rendered content (to ensure that elements are visible, and to map detected ad-identifiers to ads). As we show in Section~\ref{ssec:attacks_segmentation}, this step is non-trivial and exploitable if ad-blockers do not closely emulate the browser's DOM rendering, a complex process that varies across browsers. 
For instance, image fragmentation or spriting (see Figure~\ref{fig:adchoices-sprites}) are simple obfuscation techniques that fool Ad-Highlighter, and would engender another cat and mouse game. To avoid this, ad-blockers can directly operate on rendered images of a page, which many browsers (e.g., Chrome and Firefox) make available to extensions. 
Instead of operating on an entire rendered web page (see page-based ad-blockers below), DOM features can still be used to segment a page into regions likely to contain ads.  
For example, segmenting a page into screenshots of each \texttt{iframe} is a good starting point for detecting ads from external ad networks. The approach of Percival is also frame-based but directly relies on image frames produced during the browser's rendering process~\cite{percival}.

We consider two ways to classify frames. The first searches for ad-disclosures in rendered ads. Template-matching is insufficient due to the variability of backgrounds that ad-disclosures are overlaid on. Instead, we view this as an object-detection problem and address it with supervised ML. 
The second approach is to train a visual classifier to directly detect ad content. Hussain et al.~\cite{hussain2017automatic} report promising results for this task. Percival also relies on a lightweight deep learning model to classify frames as ad content~\cite{percival}.

\subsubsection{Page-based Perceptual Ad-blocking.}
The core idea of perceptual ad-blocking is to emulate the way humans detect ads. Element- and frame-based approaches embrace this goal to some extent, but still rely on DOM information that humans are oblivious to. Recently, Adblock Plus proposed an approach that fully emulates visual detection of online ads from rendered web content alone~\cite{sentinel}. 

In a page-based ad-blocker, segmentation is integrated into the classifier. Its core task is best viewed as an object-detection problem: given a web page screenshot, identify the location and dimension of ads. Adblock Plus trained the YOLOv3 object-detector~\cite{yolo3} on screenshots of Facebook with ads labeled using standard filter-lists. 

Once ad locations are predicted, the ad-blocker can overlay them to hide ads, or remove the underlying HTML elements (e.g., by using the \texttt{document.elementFromPoint} browser API to get the HTML element rendered at some coordinate).

%% file: attacks_new.tex

\section{Attacks on Perceptual Ad-Blocking}
\label{sec:attacks}

Given the unified architecture from Section~\ref{sec:percept_adb}, we now perform a comprehensive security analysis of the perceptual ad-blocking pipeline and describe multiple attacks targeting concrete instantiations of each of the ad-blocker's components. 
The primary focus of our analysis is to evaluate the robustness of the ad-blocker's core visual classifier, by instantiating adversarial examples for seven different and varied approaches to ad-detection (\Cref{ssec:attacks_classification}). We further demonstrate powerful attacks that exploit the ad-blocker's high-privilege actions (\Cref{ssec:attacks_action}). We conclude by describing more classical attacks that affect the segmentation step of current perceptual ad-blockers (\Cref{ssec:attacks_segmentation}), as well as potential attacks on an ad-blocker's offline data collection and training phase (\Cref{ssec:attacks_data}).

Our attacks can be mounted by different adversaries (e.g., publishers, ad-networks, or malicious third parties) to evade or detect ad-blocking and, at times, abuse the ad-blocker's high privilege level to bypass web security boundaries. These attacks, summarized in \Cref{tab:attacks}, challenge the belief that perceptual signals can tilt the arms race with publishers and ad-networks in favor of ad-blockers.
All our data, pre-trained models, and attack code are available at \url{https://github.com/ftramer/ad-versarial}.

The attacks described in this section do not violate existing laws or regulations on deceptive advertising, as the changes to the visual content of a page are imperceptible to human users. 


\subsection{Evaluation Setup}
\label{ssec:eval_setup}

\begin{table*}[!htb]
	\captionof{table}{\textbf{Evaluation of Ad-Classifiers.} For each classifier, we first evaluate on ``benign'' data collected from websites. We report false-positives (FP)---mis-classified non-ad content---and false negatives (FN)---ad-content that the classifier missed. We then give the the attack model(s) considered when \emph{evading} the classifier, the success rate, and the corresponding section.}
	\vspace{-0.75em}
	\centering
	\setlength{\tabcolsep}{3pt}
	\def\arraystretch{0.95}
	\footnotesize
	\begin{threeparttable}
		\begin{tabular}{@{}
				p{1.9cm}@{\extracolsep{4pt}}l@{\extracolsep{4pt}} l r r l@{\extracolsep{3pt}}r@{}}
			&&&\multicolumn{2}{c}{\textbf{Benign Eval.}}
			&\multicolumn{2}{c}{\textbf{Adversarial Eval.}}
			\\[-0.25em]
			\cmidrule(lr){4-5}\cmidrule(){6-7}
			\textbf{Category} & \textbf{Method} & 
			\textbf{Targets} &
			\multicolumn{1}{c}{\textbf{FP}} & \multicolumn{1}{c}{\textbf{FN}}
			& \multicolumn{1}{c}{\textbf{Attack Model for Evasion}} & \textbf{Success}
			\\
			\toprule
			\multirow{4}{1.9cm}{Element-based} 
			& Blacklist
			& AdChoices logos
			& $0/824$ & $33/41$
			& N.A.
			& \multicolumn{1}{c}{-} 
			\\
			& Avg.~hash~\cite{storey} 
			& AdChoices logos
			& $3/824$ & $3/41$
			& Add $\leq3$ empty rows/cols 
			& $100\%$
			\\
			& SIFT 
			& textual AdChoices
			& $2/824$ & $0/17$
			& $\ell_2 \leq 1.5$ 
			& $100\%$
			\\
			& OCR~\cite{storey} 
			& textual AdChoices
			& $0/824$ & $1/17$
			& $\ell_2 \leq 2.0$ 
			& $100\%$
			\\
			\midrule
			\multirow{2}{1.9cm}{Frame-based} 
			& YOLOv3 
			& AdChoices in \texttt{iframe}
			& $0/20$ & $5/29$
			& $\ell_\infty \leq {4}/{255}$ & $100\%$
			\\
			& ResNet~\cite{hussain2017automatic} 
			& ad in \texttt{iframe}
			& $0/20$ & $21/39$
			& $\ell_\infty \leq {2}/{255}$ 
			& $100\%$
			\\
			& Percival~\cite{percival} 
			& large ads in \texttt{iframe}
			& $2/7$ & $3/33$
			& $\ell_\infty \leq {2}/{255}$ 
			& $100\%$
			\\
			\midrule
			\multirow{1}{1.9cm}{Page-based}
			& YOLOv3 
			& ads visible in page screenshot
			& $2$ & $6/30$
			& Publisher: universal full-page mask ($99\%$ transparency)
			& $100\%$
			\\
			&&&&&
			Publisher: adv.~content below ads on \url{BBC.com}, $\ell_\infty \leq {3}/{255}$
			& $100\%$
			\\
			&&&&&
			Ad network: universal mask for ads on \url{BBC.com}, $\ell_\infty \leq {4}/{255}$
			& $95\%$
			\\
			\bottomrule
		\end{tabular}
	\end{threeparttable}
	\setlength{\tabcolsep}{6pt}
	\label{tab:model-accuracy}
	\vspace{-0.5em}
\end{table*}

\subsubsection{Evaluated Approaches}
We analyze a variety of techniques to instantiate the different stages of the perceptual ad-blocking pipeline. 
In particular, we evaluate \emph{seven} distinct approaches to the ad-blocker's core visual ad-classification step (see Table~\ref{tab:model-accuracy}). Three are element-based, three frame-based, and one page-based. These seven classifiers are taken from or inspired by prior work. They are:
Two computer vision algorithms used in Ad-Highlighter~\cite{storey,adhighlighter} (average hashing and OCR);  two ad classifiers, one from Hussain et al.~\cite{hussain2017automatic} and one used in Percival~\cite{percival}; a robust feature matcher, SIFT~\cite{lowe2004distinctive}; and two object detector networks---with the same YOLOv3 model~\cite{yolo3} as Sentinel~\cite{sentinel,sentinel-medium}---which we trained to detect either ad-disclosures in frames, or ads in a full web page. 

For the two object detector models we built, we explicitly separated (i.e., assigned to non-communicating authors) the tasks of (1) data-collection, design and training; and (2) development of attacks, to ensure fair evaluation results.
Our first (frame-based) model was trained to detect AdChoices logos that we overlaid in a dataset of $6{,}320$ ads collected by Hussain et al.~\cite{hussain2017automatic}.
We then classify an \texttt{iframe} as an ad, if the model detects the AdChoices logo in it.

Our second model emulates the approach of the unreleased Sentinel~\cite{sentinel, sentinel-medium} and was trained to detect ads in arbitrary news websites. This broadens Sentinel's original scope (which was limited to Facebook)---a decision we made due to difficulties in collecting sufficient training data~\cite{sentinel-medium}.
One author trained YOLOv3 to locate ads in screenshots of news websites from all G20 nations. To collect a diverse dataset of labeled ads in web screenshots, we first locates ads using a web-proxy based on filter lists, and then randomly replace ads with a larger variety of examples.
More details about this process, of independent interest, are in Appendix~\ref{apx:yolo}. A video of our model in action on five websites not seen during training is available at {\url{https://github.com/ftramer/ad-versarial/blob/master/videos}}.

\subsubsection{Evaluation Data}
We use real website data to evaluate the accuracy and robustness of the above seven ad-classifiers. We built an evaluation set from the top ten news websites in the Alexa ranking (see Table~\ref{tab:websites}). 
For each website, we extract the following data:
\begin{myenumerate}
	\item All images smaller than $50$KB in the DOM. This data is used to evaluate element-based techniques. We collect $864$ images, $41$ of which are AdChoices logos ($17/41$ logos contain the ``AdChoices'' text in addition to the icon).
	\item A screenshot of each \iframe in the DOM tree, to evaluate frame-based models. We collect $59$ frames. Of these, $39$ are ads and $29$ contain an AdChoices logo. Percival~\cite{percival} only considers images of dimension at least $100 \times 100$ px so we limit it to these.\footnote{Taking a screenshot of an \iframe is an approximation of how Chromium's rendering engine segments frames for Percival's classifier. We verified that our attacks on Percival's network work when deployed inside the Chromium browser.}
	
	\item Two screenshots per website (the front-page and an article) taken in Google Chrome on a $1920\times1080$ display.\footnote{We experimentally verified that our attacks on page-based ad-blockers are robust to changes in the user's viewport. An attacker could also explicitly incorporate multiple browsers and display sizes into its training set to create more robust attacks. Alternatively, the adversary could first detect the type of browser and viewport (properties that are easily and routinely accessed in JavaScript) and then deploy ``responsive'' attacks tailored to the user's setting.} These are used to evaluate page-based models. Each screenshot contains $1$ or $2$ fully visible ads, with $30$ ads in total.
\end{myenumerate}

For template-matching approaches (perceptual hashing and SIFT) we use the same $12$ AdChoices templates as Ad-Highlighter~\cite{adhighlighter}. 

When describing an ad-blocker's page segmentation and the corresponding markup obfuscation attacks in Section~\ref{ssec:attacks_segmentation}, we use some data collected on Facebook.com in November 2018. As Facebook continuously and aggressively adapts the obfuscation techniques it uses to target ad-blockers~\cite{ublock-forum}, the specific attacks we describe may have changed, which only goes to illustrate the ongoing arms race and need for more robust markup-less ad-blocking techniques.

\subsubsection{Accuracy and Performance of ML Classifiers}
Table~\ref{tab:model-accuracy} reports the accuracy of the seven ad-classifiers on our evaluation data. For completeness, we include a blacklist that marks any image that exactly matches one of the $12$ AdChoices logos used in Ad-Highlighter. As posited by Storey et al.~\cite{storey}, this approach is insufficient.

Note that the datasets described above are incomparable. 
Some ads are not in \texttt{iframes}, or have no ad-disclosure, ans screenshots only contain images within the current view.
Thus, the accuracy of the classifiers is also incomparable.
This does not matter, as our aim is not to find the best classifier, but to show that \emph{all} of them are insecure in the stringent attack model of visual ad-blockers.

Overall, element-based approaches have high accuracy but may suffer from some false-positives (i.e., non-ad content classified as ads) that can lead to site-breakage. The frame-based approaches are less accurate but have no false-positives. Finally, our Sentinel-like detector shows promising (albeit imperfect) results that demonstrate the possibility of ad-detection on arbitrary websites.

We measure performance of each classifier on an Intel Core i7-6700 Skylake Quad-Core 3.40GHz. While average hashing and SIFT process all images in a page in less than 4 seconds, OCR is much slower (Ad-Highlighter disables it by default). Our OCR model parses an image in $100$ ms, a $14$ second delay on some websites. The frame-based classifiers process all \texttt{iframes} in 1-7 seconds. Our page-based model processes pages downsized to $416\times416$px at $1.5$ frames-per-second (on CPU), which may suffice for ad-blocking. The authors of Percival recently demonstrated that an optimized deployment of perceptual ad-blocking with a deep learning classifier incurs only minimal overhead on page rendering ($<200$ ms).

\subsection{Attacks against Classification with Adversarial Examples}
\label{ssec:attacks_classification}

For perceptual ad-blockers that operate over images (whether on segmented elements as in Ad-Highlighter~\cite{adhighlighter}, or rendered content as in Sentinel~\cite{sentinel} or Percival~\cite{percival}), security is contingent on the robustness of the ad-blocker's visual classifier. \emph{False negatives} result in ads being shown, and \emph{false positives} cause non-ads to be blocked.

Both error types are exploitable using \emph{adversarial examples}~\cite{szegedy2013intriguing, goodfellow2014explaining}---small input perturbations that fool a classifier. Adversarial examples can be used to generate web content that fools the ad-blocker's classifier, without affecting a user's browsing experience.

In this section, we describe and evaluate four concrete types of attacks on the seven visual classifiers we consider:  (C1) \emph{adversarial ad-disclosures} that evade detection; (C2) \emph{adversarial ads} that evade detection; (C3) \emph{adversarial non-ad content} that alters the classifier's output on nearby ads; (C4) \emph{adversarial honeypots} (misclassified non-ad elements, to detect ad-blocking). Our attacks allow adversaries to evade or detect ad-blocking with (near)-100\% probability.

\paragraph{Attack Model.}
We consider adversaries that perturb web content to produce \emph{false-negatives} (to evade ad-blocking) or \emph{false-positives} (honeypots to detect ad-blocking).
Each attack targets a single classifier---but is easily extended to multiple models (see Section~\ref{sec:discussion}).

\begin{myitemize}
	\item {\em False negative.}
	To evade ad-blocking, publishers, ad networks or advertisers can perturb any web content they control, but aim to make their attacks imperceptible.
	We consider perturbations with small $\ell_2$ or $\ell_\infty$ norm (for images with pixels normalized to $[0,1]$)---a sufficient condition for imperceptibility. An exception to the above are our attacks on average hashing, which is by design invariant to small $\ell_p$ changes but highly vulnerable to other imperceptible variations. The attack model used for all evasion attacks are summarized in Table~\ref{tab:model-accuracy}.
	
	\item {\em False positive.}
	The space of non-disruptive false positive attacks is vast. We focus on one easy-to-deploy attack, that generates near-uniform rectangular blocks that blend into the page's background yet falsely trigger the ad-detector.
\end{myitemize}

We assume the publisher controls the page's HTML and CSS, but cannot access the content of ad frames. This content, including the AdChoices logo, 
is added by the ad network.


Gilmer et al.~\cite{gilmer2018motivating} argue that the typical setting of adversarial examples, where the adversary is restricted to finding imperceptible perturbations for given inputs, is often unrepresentative of actual security threats. Interestingly, the threat model for visual ad classifiers does align perfectly with this setting. The ad-blocker's adversaries want to evade its classifier for a specific input (e.g., the publisher's current web page and an advertiser's latest ad campaign), while ensuring that the users' browsing experience is unaffected.

\begin{table}[t]
	\caption{\textbf{Attack Strategies on Perceptual Ad-Blockers.} Strategies are grouped by the component that they exploit---(D)ata collection, (S)egmentation, (C)lassification, (A)ction. For each strategy, we specify which goals it can achieve, which adversaries can execute it, and which ad-blockers it applies to (fully: \CIRCLE~or partially: \LEFTcircle).\\[-2.5em]}
	\centering
	\bgroup
	\footnotesize
	\def\arraystretch{1}
	\setlength{\tabcolsep}{0.8pt}
	{
		\fontsize{7.5}{0}
		\begin{tabularx}{\columnwidth}{@{} 
				p{150pt}
				@{\extracolsep{4pt}}c@{\extracolsep{0.2pt}}c@{\extracolsep{0.2pt}}c
				@{\extracolsep{5pt}}c@{\extracolsep{0.2pt}}c@{\extracolsep{0.2pt}}c@{\extracolsep{0.2pt}}c
				@{\extracolsep{5pt}}c@{\extracolsep{0.2pt}}c@{\extracolsep{0.2pt}}c
				@{}}
			&\multicolumn{3}{c}{\textbf{Goals}}
			&\multicolumn{4}{c}{\textbf{Actors}} 
			&\multicolumn{3}{c}{\textbf{Target}}\\[-0.4em]
			\cmidrule{2-11}
			\textbf{Strategy} &  
			\rot{Evasion} & 
			\rot{Detection} & 
			\rot{Abuse} & 
			
			\rot{Publisher} & 
			\rot{Ad Network} & 
			\rot{Advertiser} & 
			\rot{Content creator} &
			
			\rot{Element-based} & 
			\rot{Frame-based} & 
			\rot{Page-based}
			\\
			\toprule
			D1: Data Training Poisoning
			& \CIRCLE & \CIRCLE & \CIRCLE
			& \CIRCLE & \CIRCLE & \CIRCLE & \CIRCLE 
			& \CIRCLE & \CIRCLE & \CIRCLE 
			\\
			\midrule
			S1: DOM Obfuscation
			& \CIRCLE & \CIRCLE & \Circle 
			& \CIRCLE & \CIRCLE & \Circle & \Circle  
			& \CIRCLE & \LEFTcircle & \Circle \\
			S2: Resource Exhaustion  (over-Segmentation)
			& \CIRCLE & \Circle & \Circle  
			& \CIRCLE & \CIRCLE & \LEFTcircle & \LEFTcircle 
			& \CIRCLE & \LEFTcircle & \Circle \\
			\midrule
			C1: Evasion with Adversarial Ad-Disclosures
			& \CIRCLE & \Circle & \Circle 
			& \Circle & \CIRCLE & \Circle & \Circle 
			& \CIRCLE & \Circle & \Circle 
			\\
			C2: Evasion with Adversarial Ads 
			& \CIRCLE & \Circle & \Circle 
			& \Circle & \CIRCLE & \CIRCLE & \Circle 
			& \Circle & \CIRCLE & \CIRCLE 
			\\
			C3: Evasion with Adversarial Content
			& \CIRCLE & \Circle & \Circle 
			& \CIRCLE & \Circle & \Circle & \CIRCLE 
			& \Circle & \Circle & \CIRCLE 
			\\
			C4: Detection with Adversarial Honeypots
			& \Circle & \CIRCLE & \Circle 
			& \CIRCLE & \Circle & \Circle & \Circle 
			& \CIRCLE & \CIRCLE  & \CIRCLE
			\\
			\midrule
			A1: Cross-Boundary Blocking 
			& \Circle & \Circle & \CIRCLE 
			& \Circle & \Circle & \CIRCLE & \CIRCLE 
			& \Circle & \Circle & \CIRCLE \\
			A2: Cross-Origin Web Requests
			& \Circle & \Circle & \CIRCLE 
			& \Circle & \Circle & \CIRCLE & \CIRCLE 
			& \CIRCLE & \Circle & \Circle \\
			\bottomrule
		\end{tabularx}
	}
	\setlength{\tabcolsep}{6pt}
	\egroup
	\label{tab:attacks}
	\vspace{-1em}
\end{table}

\subsubsection{Overview of Attack Techniques and Results}
For all seven ad-classifiers, we craft imperceptible adversarial perturbations for ad-disclosures, ads and other web content, which can be used by publishers, ad-networks, or advertisers to evade or detect ad-blocking.

\begin{figure}[t]
	\begin{minipage}{\columnwidth}
		\centering
		\footnotesize
		{
			\renewcommand{\arraystretch}{0.25}
			\setlength{\tabcolsep}{2pt}
			\belowrulesep = 0mm
			\begin{tabular}{@{} r rr r @{}}
				\multicolumn{1}{c}{\textbf{Original}} & 
				\multicolumn{1}{c}{\textbf{Avg. Hash}} & 
				\multicolumn{1}{c}{\textbf{OCR}} & 
				\multicolumn{1}{c}{\textbf{SIFT}} \\
				\toprule
				\includegraphics[height=11px,valign=T]{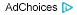}
				&
				\includegraphics[height=12.5px,valign=T]{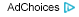}
				&
				\includegraphics[height=11px,valign=T]{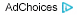} 
				&
				\includegraphics[height=11px,valign=T]{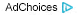}
				\\
				\includegraphics[height=11px,valign=T]{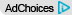}\hspace{1pt} 
				&
				\includegraphics[height=13px,valign=T]{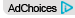}\hspace{1pt}
				&
				\includegraphics[height=11px,valign=T]{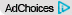}\hspace{1pt} 
				&
				\includegraphics[height=11px,valign=T]{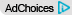}\hspace{1pt}
				\\
				\includegraphics[height=11px,valign=T]{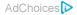}
				&
				\includegraphics[height=13px,valign=T]{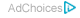}
				&
				\includegraphics[height=11px,valign=T]{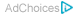}
				&
				\includegraphics[height=11px,valign=T]{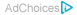}
				\\[-0.25em]
				\midrule
				\raisebox{-7pt}{%
					\setlength{\fboxsep}{0pt}%
					\setlength{\fboxrule}{0pt}%
					\fbox{\textbf{False Positives:}}%
				}\hspace{3pt}
				&
				{%
					\setlength{\fboxsep}{0pt}%
					\setlength{\fboxrule}{0.1pt}%
					\fbox{%
						\includegraphics[height=9px,valign=T]{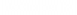}\hspace{3pt}
					}%
				}%
				&
				\includegraphics[height=10px,valign=T]{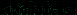}\hspace{2pt}
				&
				{%
					\setlength{\fboxsep}{0pt}%
					\setlength{\fboxrule}{0.1pt}%
					\fbox{\includegraphics[height=9px,valign=T]{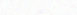}\hspace{3pt}
					}%
				}%
				\\
				\bottomrule
			\end{tabular}
		}%
		\vspace{-1.5em}
		\caption{\textbf{Adversarial Examples for Element-Based Classifiers.} These correspond to attacks (C1) and (C4) in Table~\ref{tab:attacks}.}
		\label{fig:element-based-adv-examples}
	\end{minipage}
	\\[0.5em]
	\begin{minipage}{\columnwidth}
		\centering
		\footnotesize
		\renewcommand{\arraystretch}{0.5}
		\begin{tabular}{@{} c c c @{}}
			\textbf{Original} & \textbf{False Negative} & \textbf{False Positive} \\
			\toprule
			\includegraphics[width=0.3\columnwidth,valign=M]{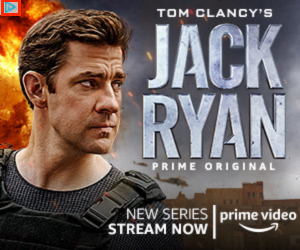}
			&
			\includegraphics[width=0.3\columnwidth,valign=M]{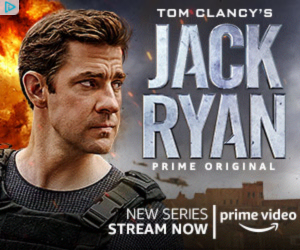}
			&
			{%
				\setlength{\fboxsep}{0pt}%
				\setlength{\fboxrule}{0.1pt}%
				\fbox{%
					\includegraphics[width=0.13\columnwidth,valign=M]{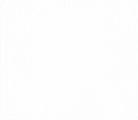}%
				}%
			}%
			\\
			\\
			\includegraphics[width=0.3\columnwidth,valign=M]{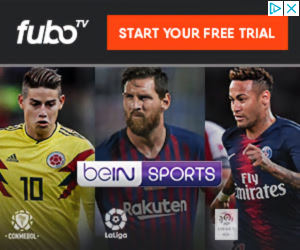}
			&
			\includegraphics[width=0.3\columnwidth,valign=M]{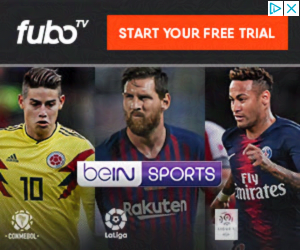}
			&
			{%
				\setlength{\fboxsep}{0pt}%
				\setlength{\fboxrule}{0.1pt}%
				\fbox{%
					\includegraphics[width=0.13\columnwidth,valign=M]{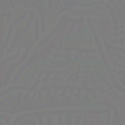}%
				}%
			}%
		\end{tabular}
		\vspace{-1em}
		\captionof{figure}{\textbf{Adversarial Examples for Frame-based Classifiers.} These are attacks (C2) and (C4) in Table~\ref{tab:attacks}. Top: Attacks on our YOLOv3 model that detects the AdChoices logo. Bottom: attacks on the ad-classifier from~\cite{hussain2017automatic} (we crafted similar adversarial examples for the classifier used in Percival~\cite{percival})\\[-2em]}
		\label{fig:frame-based-adv-examples}
	\end{minipage}
\end{figure}

Some of our classifiers can be attacked using existing techniques. For example, we show that ad-networks and publishers can use standard gradient-based attacks (see Section~\ref{ssec:adv_examples}) to create imperceptibly perturbed ads or background content that fool our two frame-based classifiers with 100\% success rates (see Figure~\ref{fig:frame-based-adv-examples}). We verify that similar attacks bypass the model used in Percival~\cite{percival}.

Attacking element-based classifiers is less straightforward, as they operate on small images (adversarial examples are presumed to be a consequence of high dimensional data~\cite{gilmer2018adversarial}), and some rely on traditional computer vision algorithms (e.g., average hashing or SIFT) for which gradient-based attacks do not apply. Nevertheless, we succeed in creating virtually invisible perturbations for the AdChoices logo, or background honeypot elements, that fool these classifiers (see Figure~\ref{fig:element-based-adv-examples}).
Our attacks on Ad-Highlighter's OCR network build upon prior work by Song and Shmatikov~\cite{song2018fooling}. 
For non-parametric algorithms such as SIFT, we propose a new generic attack using black-box optimization~\cite{salimans2017evolution, ilyas2018black} (see Section~\ref{ssec:new_attacks}), that is conceptually simpler than previous attacks~\cite{hsu2009secure}.

Our most interesting attacks are those that target page-based ad-blockers such as Sentinel~\cite{sentinel} (see Figure~\ref{fig:page-based-all}, as well as Figure~\ref{fig:bbc-apx}). Our attacks let publishers create perturbed web content to evade or detect ad-blocking, and let ad-networks perturb ads that evade ad-blocking on the multitude of websites that they are deployed in. These attacks overcome a series of novel constraints.

First, attacks on visual ML classifiers often assume that the adversary controls the full digital image fed to the classifier. This is not the case for page-based ad-blockers, whose input is a screenshot of a web document with content controlled by different actors (e.g., ad networks only control the content of ad frames, while publishers can make arbitrary website changes but cannot alter ads loaded in cross-origin \texttt{iframes}). Moreover, neither actor precisely knows what content the other actors will provide. Adversarial examples for page-based ad-blockers thus need to be encoded into the HTML elements that the adversary controls, and must be robust to variations in other page content. We solve this constraint with techniques similar to those used to make physical-world adversarial examples 
robust to random transformations~\cite{sharif2016accessorize, kurakin2016adversarial, eykholt2018physical}. We consider multiple tricks to encode a publisher's perturbations into valid HTML One attack uses CSS rules to overlay a near-transparent perturbed mask over the full page (Figure~\ref{fig:page-based-all} (b)). To detect ad-blocking, we craft an innocuous page-footer that triggers the ad-blocker (Figure~\ref{fig:page-based-all} (d)). Details on our attacks are in Section~\ref{ssec:new_attacks}.

A further challenge is the deployment of these attacks at scale, as creating perturbations for every ad and website is intractable. This challenge is exactly addressed by attacks that create \emph{universal adversarial examples}~\cite{moosavi2017universal}---\emph{single} perturbations that are crafted so as to be effective when applied to most classifier inputs. Universal perturbations were originally presented as a curious consequence of the geometry of ML classifiers~\cite{moosavi2017universal}, and their usefulness for the scalability of attacks had not yet been suggested.

Attacks on page-based ad-blockers have unique constraints, but also enable unique exploits.
Indeed, as a page-based classifier produces outputs based on a single full-page input, perturbing content controlled by the attacker can also affect the classifier's outputs on unperturbed page regions. The effectiveness of such attacks depends on the classifier. For the YOLOv3~\cite{yolo3} architecture, we show that publishers can perturb website content near ad \texttt{iframes} so as to fool the classifier into missing the actual ads (see Figure~\ref{fig:bbc-apx}).

\subsubsection{Algorithms for Adversarial Examples}
\label{ssec:new_attacks}

For some of the considered classifiers, adversarial examples for each of the attack strategies C1-C4 in Table~\ref{tab:attacks} can be constructed using existing and well-known techniques (see Section~\ref{ssec:adv_examples}). Below, we provide more details on the attack we use to target SIFT, and on the techniques we use to create robust and scalable attacks for page-based classifiers~\cite{sentinel}.

\paragraph{Black-box optimization attacks for non-parametric classifiers}
SIFT is a non-parametric algorithm (i.e., with no learned parameters). As such, the standard approach for generating adversarial examples by minimizing the model's training-loss function does not apply~\cite{szegedy2013intriguing}. To remedy this, we first formulate a near-continuous loss function $\mathcal{L}_{\textrm{SIFT}}(x+\delta)$ that acts as a proxy for SIFT's similarity measure between the perturbed image $x+\delta$ and some fixed template. The next difficulty is that this loss function is hard to differentiate, so we use black-box optimization techniques~\cite{ilyas2018black,salimans2017evolution} to minimize $\mathcal{L}_{\textrm{SIFT}}$. 

SIFT's output is a variable-sized set of keypoints, where each keypoint is a vector $v \in \R^{132}$---four positional values, and a 128-dimensional \emph{descriptor}~\cite{lowe2004distinctive}.
Let $t$ be a template with keypoint descriptors $T$. To match an image $x$ against $t$, SIFT computes descriptor vectors for $x$, denoted $\{v_1, \dots, v_m\}$. Then, for each $v_i$ it finds the distances $d_{i,1}, d_{i,2}$ to its two nearest neighbors in $T$. The keypoint $v_i$ is a match if the \emph{ratio test} ${d_{i,1}}/{d_{i,2}} < \tau$ holds (where $\tau=0.6$). 
Let $M(x, t)$ be the keypoints of $x$ that match with $t$. To evade detection, we minimize the size of $M$ via the following proxy loss:
\begin{equation}
\mathcal{L}_{\rm SIFT}(x + \delta) \coloneqq \sum\nolimits_{v_i \in M_\tau(x, t)} d_{i, 2} / d_{i, 1} \;.
\end{equation}
Minimizing $\mathcal{L}$ increases ${d_{\cdot,1}}/{d_{\cdot,2}}$ for matched keypoints until they fall below the ratio test. To create false positives, we minimize an analogous loss that sums over $v_i \notin M_\tau(x, t)$ and decreases the ratio.

\paragraph{Scalable attacks with partial input control}
When attacking page-based classifiers, we need to overcome two challenges: (1) the attacker only controls part of the page content and does not know which content other actors will add; (2) the attacks should be deployable at scale for a variety of web pages and ads. To create adversarial examples under these novel constraints, we combine {universal}~\cite{moosavi2017universal} and transformation-robust~\cite{sharif2016accessorize, kurakin2016adversarial, eykholt2018robust} attacks.

To create universal perturbations, we collect additional website screenshots: $D^{\text{train}}$ is a set of $200$ screenshots of news websites, and
$D^{\text{eval}}$ contains the $20$ screenshots collected in Section~\ref{ssec:eval_setup} (no website or ad appears in both sets). We also collect $D_{\text{BBC}}^{\text{train}}$ and $D_{\text{BBC}}^{\text{eval}}$ with $180$ and $20$ screenshots from \url{bbc.com/sports}. The training sets are used to create perturbations that work for arbitrary websites or ads. We measure attacks' success rates on the evaluation sets.

We craft a perturbation $\delta$ by minimizing $\sum_{x \in D_{*}^{\text{train}}} \mathcal{L}(x \odot \delta)$, where $x \odot \delta$ means applying the perturbation $\delta$ to a page $x$. Depending on the attack,
the perturbation is added pixel-wise to a page region that the adversary controls, 
or replaces that region with $\delta$. All that remains is the design of a suitable loss function $\mathcal{L}$.

The YOLOv3 model we trained outputs multiple $B=10{,647}$ boxes for detected ads, and retains a box $b$ if its confidence---denoted $\texttt{conf}(f(x), b) $---is larger than a threshold $\tau$. To cause ads to be undetected, we thus minimize the following loss which causes all $B$ boxes to have confidence below $\tau-\kappa$, for some slack $\kappa > 0$:
{\fontsize{9}{0}%
	\begin{equation}
	\label{eq:yolo-loss-fn}
	\mathcal{L}_{\text{YOLO}}^{\text{FN}}(x\odot\delta) \coloneqq\hspace{-2pt} \sum_{1 \leq b \leq B} \max\left(\texttt{conf}(f(x\odot\delta), b) - (\tau-\kappa), 0\right) \;,
	\end{equation}%
}%

For false-positives, i.e., a fake object prediction, we instead increase all boxes' confidence up to $\tau + \kappa$ by minimizing:
{\fontsize{9}{0}%
	\begin{equation}
	\label{eq:yolo-loss-fp}
	\mathcal{L}_{\text{YOLO}}^{\text{FP}}(x\odot\delta) \coloneqq\hspace{-2pt} \sum_{1 \leq b \leq B} \max\left(\tau+\kappa - \texttt{conf}(f(x\odot\delta), b), 0\right) \;.
	\end{equation}%
}%

\subsubsection{Evaluation of Attacks}
We now instantiate and evaluate the attack strategies C1-C4 from Table~\ref{tab:attacks} on our seven ad-classifiers

\paragraph{Attack C1: Evasion with adversarial ad-disclosures}
Figure~\ref{fig:element-based-adv-examples} shows examples of perturbed AdChoices logos that fool all element-based classifiers. An ad-network can use these to evade ad-blocking.

Average hashing is invariant to small $\ell_p$ noise, but this comes at the cost of high sensitivity to other perturbations: we evade it by adding up to $3$ transparent rows and columns to the logo. When overlaid on an ad, the rendered content is identical. 

Adversarial examples for OCR bear similarities to CAPTCHAs. As ML models can solve CAPTCHAs~\cite{bursztein2014end, ye2018captcha}, one may wonder why transcribing ad disclosures is harder. The difference lies in the stronger threat model that ad-blockers face. Indeed, CAPTCHA creators have no access to the ML models they aim to fool, and must thus craft universally hard perturbations. Attacking an ad-blocker is much easier as its internal model must be public. Moreover the ad-blocker must also prevent false positives---which CAPTCHA solvers do not need to consider---and operate under stricter real-time constraints on consumer hardware.

\paragraph{Attack C2: Evasion with adversarial ads} Ad networks can directly perturb the ads they server to evade frame or page-based ad-blockers. For frame-based classifiers, the attacks are very simple and succeed with 100\% probability (see Figure~\ref{fig:frame-based-adv-examples}). We verified that the ad-classifier used by Percival~\cite{percival} is vulnerable to similar attacks. Specifically, we create a valid HTML page containing two images---an ad and an opaque white box---which are both misclassified when the page is rendered in Percival's modified Chromium browser (see Figure~\ref{fig:percival-attack}).

For our page-based model, crafting a ``doubly-universal'' perturbation that works for all ads on all websites is hard (this is due to the model's reliance on page layout for detecting ads, see Appendix~\ref{apx:yolo} for details). Instead, we show that an ad-network can create a universal perturbation that works with 100\% success rate for all ads that it serves on a specific domain (see Figure~\ref{fig:bbc-apx}). For this attack, we minimized the $\mathcal{L}_{\text{YOLO}}^{\text{FN}}$ loss over the collected screenshots in $D_{\text{BBC}}^{\text{train}}$, by applying the same perturbation $\delta$ over all ad frames.

\begin{figure}[t]
	\begin{minipage}{\columnwidth}
		\centering
		\begin{lstlisting}[style=basic]
		<div id="overlay"></div>
		\end{lstlisting}
		\vspace{-0.7em}
		\begin{lstlisting}[style=basic,language=CSS]
		#overlay {
		background-image: url("data:image/png;base64,...");
		width: 100%; height: 100%; top: 0; left: 0;
		position: fixed; z-index: 10000; pointer-events: none;
		opacity: 0.01; }
		\end{lstlisting}
		\vspace{-1.5em}
		\caption{\textbf{Code for Attack C4-U.} An adversarial mask is tiled over the full page with a small opacity factor.\\[-2em]}
		\label{fig:overlay_css}
	\end{minipage}
\end{figure}

\begin{figure*}[t]
	\begin{minipage}{\textwidth}
		\centering
		\begin{subfigure}[t]{0.49\textwidth}
			\centering
			{%
				\setlength{\fboxsep}{0pt}%
				\setlength{\fboxrule}{1pt}%
				\fbox{%
					\includegraphics[width=\textwidth-2pt]{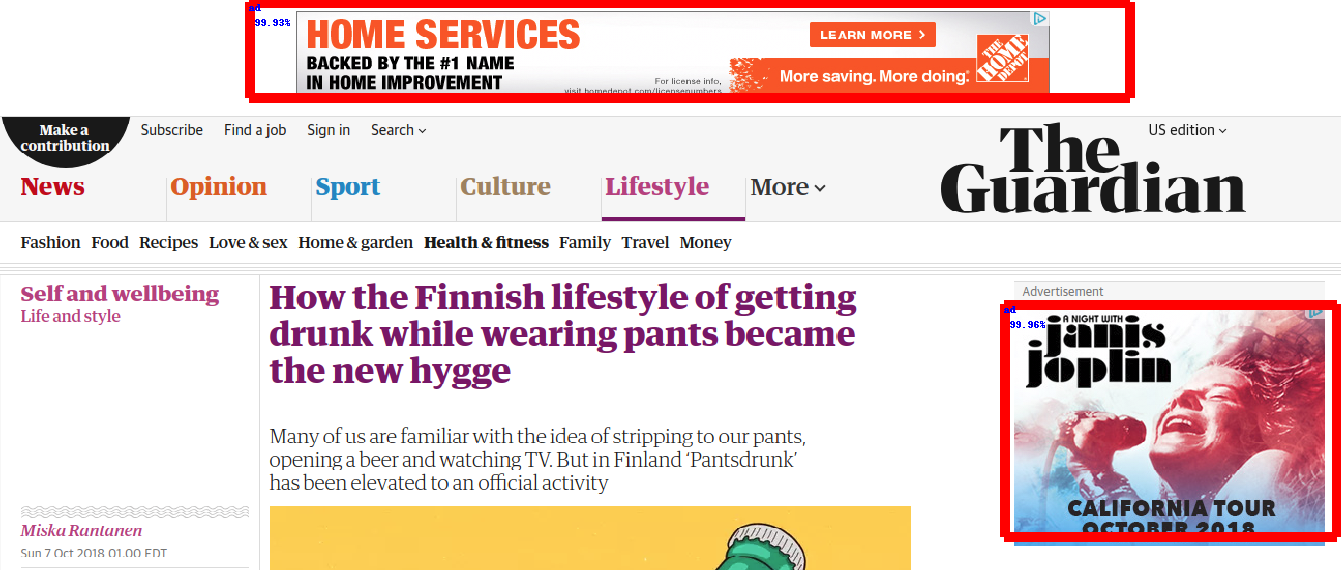}%
				}%
			}%
			\vspace{-0.4em}
			\captionof{figure}{\textbf{Original Page:} two ads are detected.}
		\end{subfigure}%
		\hfill
		\begin{subfigure}[t]{0.49\textwidth}
			\centering
			{%
				\setlength{\fboxsep}{0pt}%
				\setlength{\fboxrule}{1pt}%
				\fbox{%
					\includegraphics[width=\textwidth-2pt]{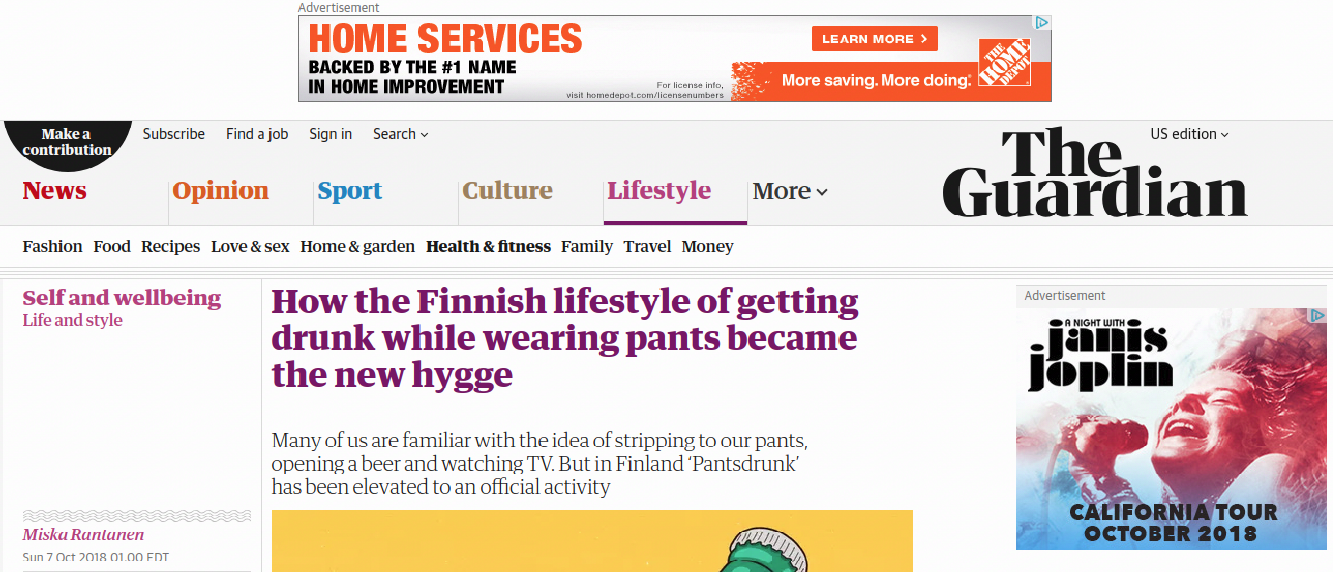}%
				}%
			}%
			\vspace{-0.4em}
			\captionof{figure}{\textbf{Attack C3 (Universal):} The publisher overlays a transparent mask over the full page to evade the ad-blocker.}
		\end{subfigure}%
		\\
		\begin{subfigure}[t]{0.49\textwidth}
			\centering
			{%
				\setlength{\fboxsep}{0pt}%
				\setlength{\fboxrule}{1pt}%
				\fbox{%
					\includegraphics[width=\textwidth-2pt]{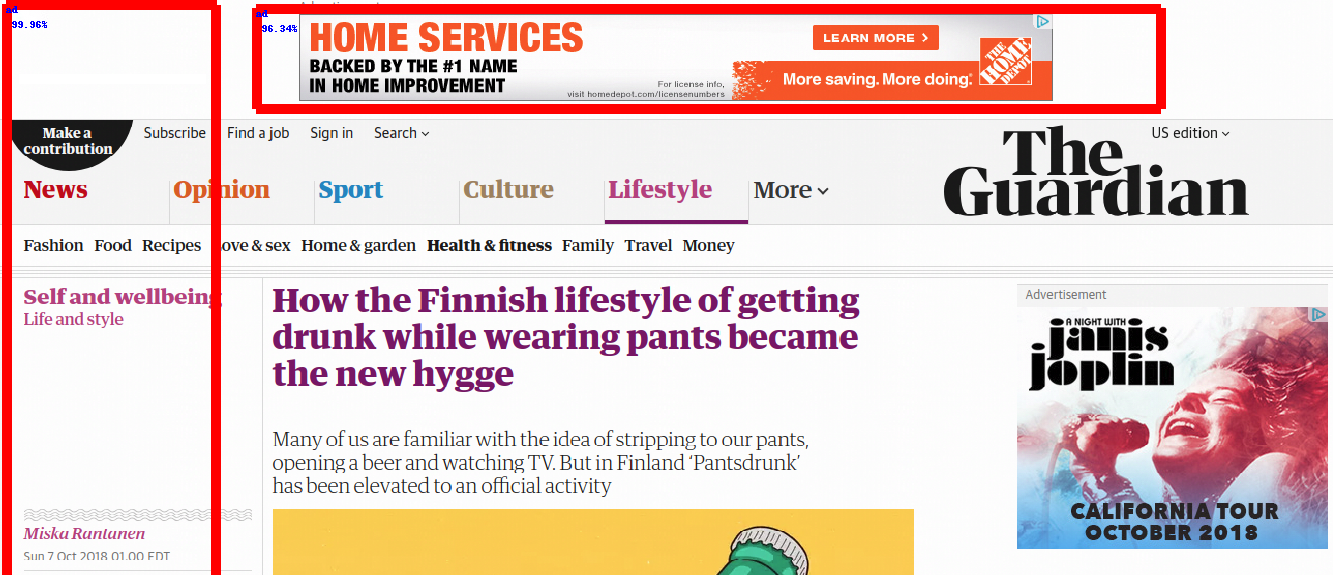}%
				}%
			}%
			\vspace{-0.4em}
			\captionof{figure}{\textbf{Attack C3 (Universal):} The publisher overlays a mask on the page to generate unreasonably large boxes and disable the ad-blocker.}
		\end{subfigure}%
		\hfill
		\begin{subfigure}[t]{0.49\textwidth}
			\centering
			{%
				\setlength{\fboxsep}{0pt}%
				\setlength{\fboxrule}{1pt}%
				\fbox{%
					\includegraphics[width=\textwidth-2pt]{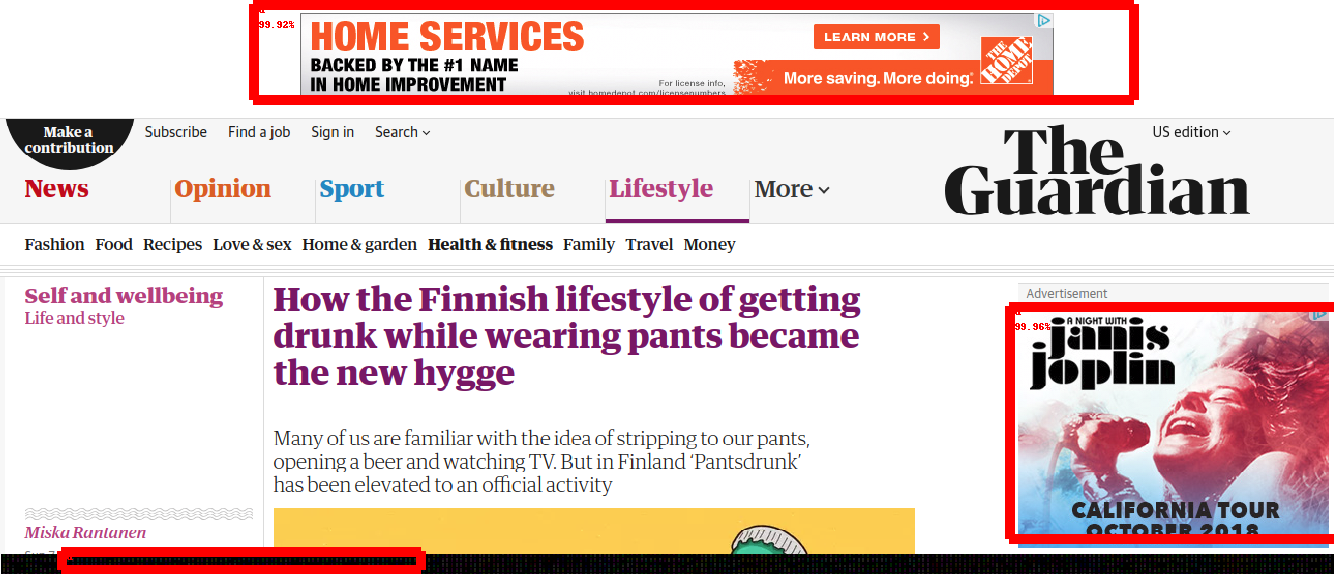}%
				}%
			}%
			\vspace{-0.4em}
			\captionof{figure}{\textbf{Attack C4 (Universal):} The publisher adds an opaque footer to detect an ad-blockers that blocks the honeypot element (bottom-left).}
		\end{subfigure}
		\vspace{-1em}
		\caption{\textbf{Universal Adversarial Examples for Page-Based Ad-Blockers.} Displays examples of universal evasion attacks (C3) and detection attacks (C4) on a page from \url{theguardian.com}. Best viewed with 2x zoom in.}
		\label{fig:page-based-all}
		\vspace{-0.5em}
	\end{minipage}
\end{figure*}

\paragraph{Attack C3: Evasion with adversarial content}
These attacks apply to page-based ad-blockers and allow publishers to evade ad-blocking while only perturbing HTML elements that they control (which crucially does not include the content of ad-frames). We show that a publisher can actually perturb the full screenshot image fed into the classifier using CSS techniques. The HTML perturbation is a \emph{near-transparent mask}, that is overlaid on the entire web page (see Figure~\ref{fig:overlay_css}). The CSS properties \texttt{z-index} and \texttt{pointer-events} are used to display the mask over all other web content, but allow underlying elements to still be accessed and clicked normally.

Adding a mask over the full image is prohibitive, as the mask would be large and tied to a fixed resolution. We thus build a smaller mask and \emph{tile} it over the full page. We generate a universal adversarial mask $\delta$ of $20$KB by minimizing $\mathcal{L}_{\text{YOLO}}^{\text{FN}}$ over $D^{\text{train}}$. The overlaid mask evades detection of all ads in our evaluation set (see Figure~\ref{fig:page-based-all}, (b)). This attack can be deployed by \emph{any} publisher, to evade all ads. The perturbation mask is robust to scrolling and viewport changes when tested in Google Chrome.

Figure~\ref{fig:page-based-all} (c) shows a similar attack that \emph{overloads} the ad-blocker. The transparent mask is crafted to minimize $\mathcal{L}_{\text{YOLO}}^{\text{FN}}$ over $D^{\text{train}}$, and creates many incorrect predictions that coerce the ad-blocker into abdicating or breaking the site. On all websites, the mask causes the model to detect abnormally large ads or fail to detect real ads.

These attacks are powerful and can be re-used by any publisher. Yet, ad-blockers might try to detect certain CSS tricks and disable them. We thus also propose stealthier attacks tuned to a single domain. For pages on \url{BBC.com}, we create a small perturbation ($40\times1020$ px) that is applied to the white background right below an ad frame (see Figure~\ref{fig:bbc-apx}(b)) and that universally applies to all pages from that publisher that use a similar layout.

\paragraph{Attack C4: Detection with adversarial honeypots}
To detect ad-blocking, publishers can use honeypots that falsely trigger ad-blockers~\cite{zhu2018measuring}. The false positives in Figures~\ref{fig:element-based-adv-examples}-\ref{fig:frame-based-adv-examples} are innocuous elements that are falsely classified as ads or ad-disclosures. For OCR and the model of Hussain et al.~\cite{hussain2017automatic}, generating near-opaque black elements worked best.
As average hashing is invariant to changes in image intensity, creating false positives for it is trivial.

For page-based ad-blockers, our first attack embeds a perturbation into a small page footer (see Figure~\ref{fig:page-based-all} (d)). The footer causes false predictions for $19/20$ pages in our evaluation set, and is robust to a user scrolling over the page. Figure~\ref{fig:bbc-apx} (c) shows a stealthier attack---tailored to bbc.com---that hides a honeypot in the page header and has 100\% success rate across pages from that publisher.

\subsection{Attacks against Ad-Blocker Actions}
\label{ssec:attacks_action}

Ad-blockers usually run at a higher privilege level than any web page. They are generally not affected by the same-origin policy and can read and write any part of any web page that the user visits.

The main privileged action taken by an ad-blocker is altering of web content. Attackers exploit this action when using honeypots to detect ad-blockers. But triggering ad-blocker actions can have more pernicious effects. Below, we describe two attacks that can be deployed by \emph{arbitrary content creators} (e.g., a Facebook user) to trigger malicious ad-blocker actions in other users' browsers. 

\paragraph{Attack A1: Cross-boundary blocking}
In this attack (see Figure~\ref{fig:tom-jerry}) a malicious user (Jerry) uploads adversarial content that triggers a Sentinel-like ad-blocker into marking content of another user (Tom) as and ad. This ``cross-boundary blocking attack'' hijacks the ad-blocker's elevated privilege to bypass web security boundaries.

To mount the attack, we optimally perturb Jerry's content so as to maximize the model's confidence in a box that covers Tom's content.
The attack works because object-detector models such as YOLOv3~\cite{yolo3} predict bounding boxes by taking into account the \emph{full} input image---a \emph{design feature} which increases accuracy and speed~\cite{yolo1}. As a result, adversarial content can affect bounding boxes in arbitrary image regions. Our attack reveals an inherent vulnerability of \emph{any} object detector applied to web content---wherein the model's segmentation misaligns with web-security boundaries.

\paragraph{Attack A2: Cross-origin web requests}
In addition to searching for the ``Sponsored'' text on Facebook, Ad-Highlighter~\cite{adhighlighter} uses the fact that the ad-disclosure contains a link to Facebook's ad-policy page as an additional signal. Specifically, Ad-Highlighter parses the DOM in search for links containing the text ``Sponsored'' and determines whether the link leads to Facebook's ad statement page by simulating a user-click on the link and following any redirects. \footnote{Ad-Highlighter simulates clicks because Facebook used to resolve links server-side (the ad-disclosure used to link to \url{www.facebook.com/\#}). Facebook recently changed its obfuscation of the link in post captions. It now uses an empty \texttt{<a>} tag that is populated using JavaScript during the click event. This change fools Ad-Highlighter and still requires an ad-blocker to simulate a potentially dangerous click to uncover the link.}

These techniques are dangerous and enable serious vulnerabilities (e.g., CSRF~\cite{pellegrino2017deemon}, DDoS~\cite{pellegrino2015cashing} or click-fraud~\cite{christin2010dissecting}) with consequences extending beyond ad-blocking. Clicking links on a user's behalf is a highly privileged action, which can thus be exploited by \emph{any party that can add links in a page}, which can include arbitrary website users. To illustrate the dangers of behavioral ad-blocking, we create a regular Facebook post with an URL to a web page with title ``Sponsored''. Facebook converts this URL into a link which Ad-Highlighter clicks on. \emph{Albeit sound, this attack luckily and coincidentally fails due to Facebook's Link Shim}, that inspects clicked links before redirecting the user. Ad-Highlighter fails to follow this particular redirection thus inadvertently preventing the attack. Yet, this also means that Facebook could use the same layer of indirection for their ``Sponsored'' link. If the behavioral ad-blocking idea were to be extended to disclosure cues on other websites (e.g., the AdChoices logo), such attacks would also be easily mounted. Pre-filtering inputs passed to a behavioral layer does not help. Either the filter is perfect, in which case no extra step is required---or its false positives can be exploited to trigger the behavioral component.

\subsection{Attacks against Page Segmentation}
\label{ssec:attacks_segmentation}

In this Section, we describe attacks targeting the ad-blocker's page segmentation logic, in an effort to evade the ad-blocker or exhaust its resources. These attacks use standard Web techniques (e.g., HTML obfuscation) and are already applied in an ongoing arms race between Facebook and uBlock~\cite{ublock-forum}. We argue that to escape the arms race caused by these segmentation attacks, perceptual ad-blockers have to operate over \emph{rendered} web-content (i.e., frame or page-based approaches), which in turn increases the attack surface for adversarial examples on the ad-blocker's visual classifier.

\begin{figure}[t]
	\begin{minipage}{\columnwidth}
		\centering
		\begin{minipage}{0.56\columnwidth}
			\begin{lstlisting}[style=basic,language=HTMLSpecial,xleftmargin={0.25cm},xrightmargin=0cm]
			<a><span>
			<span class="c1">Sp</span>
			<span class="c2">S</span>
			<span class="c1">on</span>
			<span class="c2">S</span>
			<span class="c1">so</span>
			<span class="c2">S</span>
			<span class="c1">red</span>
			<span class="c2">S</span>
			</span></a>
			\end{lstlisting}
			\vspace{-0.8em}
			\begin{lstlisting}[style=basic,language=CSS,xleftmargin={0.25cm},xrightmargin=0cm]
			.c2 { font-size: 0; }
			\end{lstlisting}
		\end{minipage}
		\hfill
		\begin{minipage}{0.42\columnwidth}
			\includegraphics[width=\textwidth]{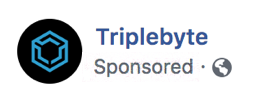}
		\end{minipage}
		\vspace{-0.8em}
		\caption{\textbf{CSS Obfuscation on Facebook.} (Left) HTML and CSS that render Facebook's ``Sponsored'' caption. (Right) A proof-of-concept where the ad-disclosure is an adversarial image that Ad-Highlighter's OCR decodes as ``8parisared''.}
		\label{fig:fb_ocr}
	\end{minipage}
	\\[1em]
	\begin{minipage}{\columnwidth}
		\centering
		\begin{tabular}{@{}c c@{}}
			{%
				\setlength{\fboxsep}{1pt}%
				\setlength{\fboxrule}{0.1pt}%
				\fbox{%
					\includegraphics[height=12px,valign=M]{figures/element_based/ac-topleft-sprite.png}%
				}%
			}%
			&
			{%
				\setlength{\fboxsep}{2pt}%
				\setlength{\fboxrule}{0.1pt}%
				\fbox{%
					\includegraphics[width=40px,valign=M]{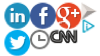}%
				}%
			}%
		\end{tabular}
		\vspace{-0.75em}
		\caption{\textbf{Image Sprites of the AdChoices Logo.} \emph{Image-sprites} are sets of images stored in a single file, and segmented using CSS rules. For example, the left sprite allows to smoothly switch from the icon to the full logo on hover. The right sprite is used by cnn.com to load a variety of logos used on the page in a single request.\\[-2.2em]}
		\label{fig:adchoices-sprites}	
	\end{minipage}
\end{figure}

\paragraph{Attack S1: DOM obfuscation}
These attacks aim to fool the ad-blocker into feeding ambiguous inputs to its classifier. 
They exploit some of the same limitations that affect traditional filter lists, and can also be applied to element-based ad-blockers that rely on computer-vision classifiers, such as Ad-Highlighter.

DOM obfuscation is exemplified by Facebook's continuous efforts to regularly alter the HTML code of its ``Sponsored'' caption (see Figure~\ref{fig:fb_ocr}). Facebook deploys a variety of CSS tricks to obfuscate the caption, and simultaneously embeds hidden ad-disclosure honeypots within regular user posts in an effort to deliberately cause site-breakage for ad-block users. Facebook's obfuscation attempts routinely fool uBlock~\cite{ublock-forum} as well as Ad-Highlighter. 

If ad-blockers adopt computer-vision techniques as in Ad-Highlighter, DOM obfuscation attacks still apply if ad-blockers assume a direct correspondence between elements in the DOM and their visual representation when rendered. For example, Ad-Highlighter assumes that all \img tags in the DOM are shown as is, thereby ignoring potentially complex CSS transformations applied when rendering HTML. This can cause the downstream classifier to process images with unexpected properties.

Ad networks already use CSS rules that significantly alter rendered ad-disclosures. \Cref{fig:adchoices-sprites} shows two AdChoices logos found on \url{cnn.com}. These are image-sprites---multiple images included in a single file to minimize HTTP requests---that are cropped using CSS to display only a single logo at a time. Image-sprites highlight an exploitable blind-spot in element-based perceptual ad-blockers---e.g., the logos in \Cref{fig:adchoices-sprites} fool Ad-Highlighter~\cite{adhighlighter}. 
Images can also be fragmented into multiple elements. The ad-blocker then has to stitch them together to correctly recognize the image (e.g., Google's AdChoices logo consists of two separate SVG tags).

Finally, the rules used by ad-blockers to link ad-disclosures back to the corresponding ad frame can also be targeted. For example, on pages with an integrated ad network, such as Facebook, the publisher could place ad-disclosures (i.e., ``Sponsored'' links) and ads at arbitrary places in the DOM and re-position them using CSS.

Frame-based and page-based ad-blockers bypass all these issues by operating on already-rendered content.

\paragraph{Attack S2: Over-segmentation}
Here the publisher injects a large number of elements into the DOM (say, by generating dummy images in JavaScript) to overwhelm an ad-blocker's classifier with inputs and exhaust its resources. In response, ad-blockers would have to aggressively filter DOM elements---with the risk of these filters' blind spots being exploited to evade or detect ad-blocking. The viability of this attack may seem unclear, as users might blame publishers for high page-load latency resulting from an overloaded ad-blocker. Yet, Facebook's efforts to cause site-breakage by embedding ad-disclosure honeypots within all regular user posts demonstrates that some ad networks may result to such tactics.

\subsection{Attacks against Training}
\label{ssec:attacks_data}

For classifiers that are trained on labeled images, the data collection and training phase can be vulnerable to \emph{data poisoning attacks} (D1)---especially when crowdsourced as with Sentinel~\cite{sentinel}. We describe these attacks for completeness, but refrain from a detailed evaluation as the test-time attacks described in Sections~\ref{ssec:attacks_classification} through~\ref{ssec:attacks_segmentation} are conceptually more interesting and more broadly applicable.

In these attacks, the adversary joins the crowdsourced data collection to submit maliciously crafted images that adversely influence the training process. For example, malicious training data can contain \emph{visual backdoors}~\cite{chen2017targeted}, which are later used to evade the ad-blocker.  The ad-blocker developer cannot tell if a client is contributing real data for training or malicious samples. 
Similar attacks against crowdsourced filter lists such as Easylist are theoretically possible. A malicious user could propose changes to filter lists that degrade their utility. However, new filters are easily interpreted and vetted before inclusion---a property not shared by visual classifiers. 

Sentinel's crowdsourced data collection of users' Facebook feeds also raises serious privacy concerns, as a deployed model might leak parts of its training data~\cite{fredrikson2015model,shokri2017membership}.

%% file: discussion.tex
\section{Discussion}
\label{sec:discussion}
We have presented multiple attacks to evade, detect and abuse recently proposed and deployed perceptual ad-blockers. We now provide an in-depth analysis of our results. 

\subsection{A New Arms Race} Our results indicate that perceptual ad-blocking will either perpetuate the arms race of filter lists, or replace it with an arms race around adversarial examples. Where perceptual ad-blockers that rely heavily on page markup (e.g., as in uBlock~\cite{ublock} or Ad-Highlighter~\cite{adhighlighter}) remain vulnerable to continuous markup obfuscation~\cite{ublock-forum}, visual classification of rendered web content (as in Sentinel~\cite{sentinel} or Percival~\cite{percival}) inherits a crucial weakness of current visual classifiers---adversarial examples~\cite{szegedy2013intriguing,goodfellow2014explaining}.

The past years have seen considerable work towards mitigating the threat of adversarial examples. Yet, defenses are either broken by improved attacks~\cite{athalye2018obfuscated,carlini2016towards}, or limited to restricted adversaries~\cite{madry2017towards,kolter2017provable,raghunathan2018certified,tramer2018ensemble, chou2018sentinet}. Even if ad-block developers proactively detect adversarial perturbations and blacklist them (e.g., using adversarial training~\cite{szegedy2013intriguing, madry2017towards} to fine-tune their classifier), adversaries can simply regenerate new attacks (or use slightly different perturbations~\cite{sharma2017attacking}).


\subsection{Strategic Advantage of Adversaries and Lack of Defenses} Our attacks with adversarial examples are not a \emph{quid pro quo} step in this new arms race, but indicate a pessimistic outcome for perceptual ad-blocking. 
Indeed, these ad-blockers operate in essentially \emph{the worst threat model for visual classifiers}. Their adversaries have access to the ad-blockers' code and prepare offline digital adversarial examples to trigger both false-negatives and false-positives in the ad-blocker's online (and time constrained) decision making.

Even if ad-blockers obfuscate their code, black-box attacks~\cite{ilyas2018black} or model stealing~\cite{tramer2016stealing,papernot2016practical} still apply. Randomizing predictions or deploying multiple classifiers is also ineffective~\cite{athalye2018obfuscated, he2017adversarial}. For example, some of the adversarial examples in Figure~\ref{fig:element-based-adv-examples} work for both OCR and SIFT despite being targeted at a single one of these classifiers.

The severity of the above threat model is apparent when considering existing defenses to adversarial examples. For instance, adversarial training~\cite{szegedy2013intriguing,madry2017towards} assumes restricted adversaries (e.g., limited to $\ell_\infty$ perturbations), and breaks under other attacks~\cite{engstrom2017rotation, sharma2017attacking,tramer2019adversarial}. Robustness to adversarial false positives (or ``garbage examples''~\cite{goodfellow2014explaining}) is even harder. Even if ad-blockers proactively re-train on adversarial examples deployed by publishers and ad-networks, training has a much higher cost than the attack generation and is unlikely to generalize well to new perturbations~\cite{schmidt2018adversarially}.
\emph{Detecting} adversarial examples~\cite{grosse2017statistical,metzen2017detecting}  (also an unsolved problem~\cite{carlini2017adversarial}) is insufficient as Ad-blockers face both adversarial false-positives and false-negatives, so merely detecting an attack does not aid in decision-making.
A few recently proposed defenses achieve promising results in some restricted threat models, e.g., black-box attacks~\cite{chen2019stateful} or physically-realizable attacks~\cite{chou2018sentinet}. These defenses are currently inapplicable in the threat model of perceptual ad-blocking, but might ultimately reveal new insights for building more robust models.

Our attacks also apply if perceptual ad-blocking is used as a complement to filter lists rather than as a standalone approach. Ad-blockers that combine both types of techniques are vulnerable to attacks targeting either.  If perceptual ad-blocking is only used passively (e.g., to aid in the maintenance of filter lists, by logging potential ads that filter lists miss), the ad-blocker's adversaries still have incentive to attack to delay the detection of new ads.

This stringent threat model above also applies to ML-based ad-blockers that use URL and DOM features~\cite{gugelmann2015automated, bhagavatula2014leveraging, iqbal2018adgraph}, which have not been evaluated against \emph{adaptive white-box} attacks.



\subsection{Beyond the Web and Vision.} The use of sensory signals for ad-blocking has been considered outside the Web, e.g., AdblockRadio detects ads in radio streams using neural networks~\cite{adblockradio}. Emerging technologies such as virtual reality~\cite{adblockVR}, voice assistants~\cite{alexaAds} and smart TVs~\cite{adblocktv} are posited to become platforms for large-scale targeted advertising, and perceptual ad-blockers might emerge in those domains as well.

The threats described in this paper---and adversarial examples in particular---are likely to also affect perceptual ad-blockers that operate outside the vision domain. To illustrate, we take a closer look at AdblockRadio, a radio client that continuously classifies short audio segments as speech, music or ads based on spectral characteristics. When ads are detected, the radio lowers the volume or switches stations. Radio ad-blockers face a different threat model than on the Web. All content, including ads, is served as raw audio from a single origin, so filter lists are useless. The publisher cannot run any client-side code, so ad-block detection is also impossible. Yet, the threat of adversarial examples does apply. Indeed, we show that by adding near-inaudible\footnote{The perturbed audio stream has a signal-to-noise ratio of $37$ dB.} noise to the ad content in AdblockRadio's demo podcast, the perturbed audio stream evades ad detection.

\begin{figure}[t]
	\includegraphics[width=\columnwidth]{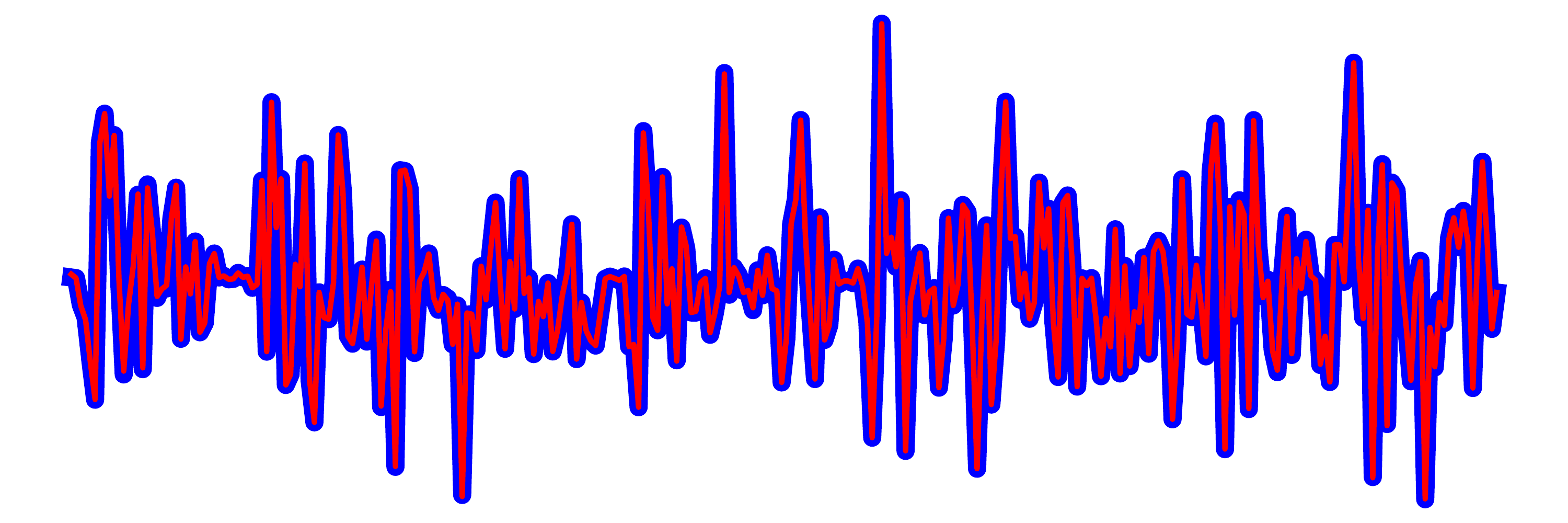}
	
	\vspace{-0.5em}
	\caption{\textbf{Original and Adversarial Audio Waveforms.} Shows a ten second segment of an ad audio waveform (thick blue) overlaid with its adversarial perturbation (thin red).}
	\label{fig:audio}
	\vspace{-1em}
\end{figure}

Concretely, AdblockRadio takes as input a raw audio stream, computes the Mel-frequency cepstral coefficients (MFCCs), and splits them into non-overlapping windows of $4$ seconds. Each segment is fed into a standard feed-forward classifier that predicts whether the segment corresponds to music, speech, or an ad.  A post-processing phase merges all consecutive segments of a same class, and removes ad-segments. As the whole prediction pipeline is differentiable, crafting adversarial examples is straightforward: we use projected gradient descent (in the $l_\infty$-norm) to modify the raw ad audio segments so as to minimize the classifier's confidence in the ad class. The resulting audio stream fully bypasses AdblockRadio's ad detection. An ad segment in the original and adversarial audio waveforms is displayed in Figure~\ref{fig:audio}.

\vspace{-0.25em}
  

%% file: relwork.tex
\section{Related Work}
\label{sec:relwork}

Our work bridges two areas of computer security research---studies of the online ad-ecosystem and associated ad-blocking arms race, and adversarial examples for ML models.

\paragraph{Behavioral advertising.}
A 2015 study found that $22\%$ of web users use ad-blockers, mainly due to intrusive behavior~\cite{Pujol:2015:AUA:2815675.2815705, ur2012smart, singh2009blocking, kontaxis2015tracking}. The use of ad-disclosures---which some perceptual ad-blockers rely on---is rising. On the Alexa top 500, the fraction of ads with an AdChoices logo has grown from $10\%$ to $60\%$ in five years~\cite{hernandez2011tracking, storey}. Yet, less than $27\%$ of users understand the logo's meaning~\cite{ur2012smart, leon2012online}.

\paragraph{Ad-blocking.}
Limitations of filter lists are well-studied~\cite{malloy2016ad, wills2016ad, vastel2018filters}. Many new ad-blocker designs (e.g., \cite{gugelmann2015automated, bhagavatula2014leveraging, iqbal2018adgraph}) replace hard-coded rules with ML models trained on similar features (e.g., markup~\cite{crites2004automatic} or URLs~\cite{krammer2008effective}). Many of these works limit their security analysis to \emph{non-adaptive} attacks. Ours is the first to rigorously evaluate ML-based ad-blockers.

Ad-block detection has spawned an arms race around anti-ad-blocking scripts~\cite{mughees2016first, mughees2017detecting, nithyanand2016adblocking}. Iqbal et al.~\cite{iqbal2017ad} and Zhu et al.~\cite{zhu2018measuring} detect anti-ad-blocking using code analysis and differential-testing. 
Storey et al.~\cite{storey} build \emph{stealthy} ad-blockers that aim to hide from client-side scripts, a challenging task in current browsers (see Appendix~\ref{apx:detection}).

\paragraph{Adversarial examples.}
Our work is the first to apply adversarial examples in a real-world web-security context. Prior work attacked image classifiers~\cite{szegedy2013intriguing,goodfellow2014explaining,papernot2016limitations,carlini2016towards}, malware~\cite{grosse2016adversarial}, speech recognition~\cite{carlini2018audio} and others. We make use of white-box attacks on visual classifiers~\cite{madry2017towards, carlini2016towards}, sequential models~\cite{song2018fooling,carlini2018audio} and object detectors~\cite{eykholt2018physical}. We show that \emph{black-box} attacks~\cite{ilyas2018black} are a generic alternative to prior attacks on SIFT~\cite{hsu2009secure}.

Attacking page-based ad-blockers introduce novel challenges. Perturbing HTML bares similarities to discrete domain attacks, e.g., PDF malware detection~\cite{laskov2014practical}. 
The ad-blocker's inputs can also be controlled by multiple entities, a constraint reminiscent of those that arise in \emph{physical-world} attacks~\cite{kurakin2016scale, sharif2016accessorize, eykholt2018robust, athalye2018synthesizing, eykholt2018physical}.

Preventing adversarial examples is an open problem. Adversarial training is a viable strategy~\cite{goodfellow2014explaining,kurakin2016scale,madry2017towards,tramer2018ensemble}, but considers a less stringent threat model than perceptual ad-blockers.

%% file: conclusion.tex
\section{Conclusion}

We have presented a comprehensive security evaluation of perceptual ad-blocking. To understand the design space of these recently deployed systems, we have derived a unified architecture that incorporates and extends prior work. Our analysis of this architecture has revealed multiple vulnerabilities at every stage of the visual ad-classification pipeline. We have shown that unless perceptual ad-blockers operate over rendered web content, the arms race around page markup obfuscation will likely carry on. Conversely, we have demonstrated that current visual ad-classifiers
are inherently vulnerable to adversarial examples---the first application of these attacks to web-security. We have shown how to craft near-imperceptible perturbation for ads, ad-disclosures, and native content, in order to evade or detect ad-blocking with seven different classifiers. Finally, we have discovered a powerful attack on page-based ad-blockers, wherein a malicious user fools the model into blocking content supposedly protected by web-security boundaries.

Our aim was to highlight the fundamental vulnerabilities that perceptual ad-blockers inherit from existing image classifiers. As long as defenses to adversarial examples are elusive, perceptual ad-blockers will be dragged into a new arms race in which they start from a precariously disadvantaged position---given the stringent threat model that they must survive.

\section*{Acknowledgments}
This work was partially supported by NSF, ONR, the Simons Foundation,
a Google faculty fellowship, the Swiss National Science Foundation 
(SNSF project P1SKP2\_178149), and the German Federal Ministry
of Education and Research (BMBF) through funding for the
CISPA-Stanford Center for Cybersecurity (FKZ: 13N1S0762).

%% file: appendix_detection.tex
\section{The Ad-block Detection Arms Race}
\label{apx:detection}

Many publishers actively detect the presence of ad-blockers~\cite{mughees2016first,nithyanand2016adblocking,iqbal2017ad} and take actions ranging from user warnings to service disabling for ad-block users. Ad-block detection operates among three main axes~\cite{storey}: (1) detecting absence of known ads; 
(2) injecting ``honeypots'' and detecting that they are mistakenly blocked, and (3) detecting ad-blocking code through side-channels (e.g., timing).

Perceptual ad-blockers cannot be detected server-side as they do not alter any web requests. To remain stealthy, a perceptual ad-blocker thus only needs to fool publisher JavaScript code into observing an unmodified DOM~\cite{storey}. This challenge is surmountable for native in-browser ad-blockers, as these can simply modify the user's view without affecting the DOM. 
Yet, the main ad-blockers today are browser extensions, which do not have such high privilege levels and share the same JavaScript API as client scripts. Storey et al.~\cite{storey} suggest the following arms race for a stealthy ad-blocker:

\begin{myenumerate}
	\item The ad-blocker modifies the DOM to block or mask detected ads and honeypots. It then \emph{overwrites the JavaScript DOM traversal API} (e.g., with JavaScript \emph{proxies}) so that the publisher's code sees the original DOM.
	
	\item The publisher inspects changes to global APIs by using the \texttt{toString()} method to unveil changes on the function.\footnote{Even proxied functions can be distinguished from their native counterparts: \url{https://bugs.chromium.org/p/v8/issues/detail?id=7484}.}
	
	\item The ad-blocker overwrites the universal \texttt{toString()} method used by all JavaScript functions, so that it always returns the same value as for a non-blocked website. 
\end{myenumerate}

We argue that this is not the end of the arms race. We sketch three strategies to detect or reverse the above ad-blocker modifications. 
Preventing the attacks below requires the ad-blocker to emulate a much larger set of JavaScript APIs, parts-of-which appear inaccessible to browser extensions.

\begin{myenumerate}
	\item \textbf{Borrowing native functions.} A publisher creates an \texttt{iframe}, which gets a new JavaScript environment, and extracts a ``fresh'' native function (e.g., \texttt{toString}) from it to unveil changes. In turn, the ad-blocker has to intercept
	all \texttt{iframe} creations and re-apply the same changes.
	
	\item \textbf{Detecting non-native functions.} The \texttt{toString} method is native (i.e., implemented by the browser). Some properties differ between native and non-native functions and do not appear to be mockable (e.g., setting a native function's \texttt{arguments} property raises an error whereas this property can be set for JavaScript functions).\footnote{It might be possible to circumvent this issue with a Proxy. Yet, we found that function Proxies can be distinguished from native functions in Google Chrome via the error message of a \texttt{postMessage} call---this might be mockable too, but vastly expands the portion of the JavaScript API to cover.}
	
	\item \textbf{Timing.} If the above attacks are solved by emulating large portions of native JavaScript, the performance overhead may lead to a strong timing side-channel.
\end{myenumerate}

%% file: appendix_yolo.tex

\section{Training a Page-Based Ad-Blocker}
\label{apx:yolo}

As the trained neural network of Sentinel~\cite{sentinel} is not available for an evaluation, we trained one for the analysis of \Cref{sec:attacks}. We used the same architecture as Sentinel, i.e., YOLO (v3)~\cite{yolo1, yolo2, yolo3}.

\subsection{Data Collection}

YOLO is an object detection network. Given an image, it returns a set of bounding boxes for each detected object. To train and evaluate YOLO, we created a dataset of labeled web page screenshots where each label encodes coordinates and dimensions of an ad on the page. We created the dataset with an ad-hoc automated system that operates in two steps. First, given a URL, it retrieves the web page and identifies the position of ads in the page using filter lists of traditional ad-blockers. Then, our system generates a web page template where ads are replaced with placeholder boxes. The concept of web page templates is convenient as it enables us to create multiple screenshots from the same web page with different ads, a form of data-augmentation. Second, from each web page template, we derive a number of images by placing ads on the template. 

\paragraph{Web pages.} We acquired web pages by retrieving the URLs of the top 30 news websites of each of the G20 nations
listed in \url{allyoucanread.com}. For each news site, we searched for the RSS feed URLs and discarded sites with no RSS feeds. The total number of RSS feed URLs is 143. We visited each RSS feed URL daily and fetched the URLs to the daily news. 

\paragraph{Template generation.} Given a URL of a news article, we generate a page template using a modified HTTP proxy that matches incoming HTTP requests against traditional ad-blocker filter lists, i.e., Easylist~\cite{easylist} and Ghostery~\cite{ghostery}. The proxy replaces ad contents with monochrome boxes using a unique color for each ad. These boxes are placeholders that we use to insert new ads. We manually inspected all templates generated during this step to remove pages with a broken layout (caused by filter lists' false positives) or pages whose ads are still visible (caused by filter lists' false negatives). 

\paragraph{Image generation.} From each page template, we generate multiple images by replacing placeholder boxes with ads. We select ads from the dataset of Hussain et al.~\cite{hussain2017automatic}. This dataset contains about 64K images of ads of variable sizes and ratios. We complemented the dataset with 136 ads we retrieved online. To insert pictures inside a template, we follow four strategies: 
\begin{myenumerate}
	\item 
	We directly replace the placeholder with an ad; 
	
	\item 
	We replace the placeholder with an ad, and we also include
	an AdChoices logo in the top right corner of the ad;
	
	\item 
	We augment templates without placeholders by adding a large ad popup in the page. The page is darkened to highlight the ad;
	
	\item 
	We insert ads as background of the website, that fully cover the left- and right-hand margins of the page.
\end{myenumerate}
When inserting an ad, we select an image with a similar aspect ratio. When we cannot find an exact match, we resize the image using Seam Carving~\cite{avidan2007seam}, a content-aware image resizing algorithm that minimizes image distortion. To avoid overfitting during training, we limited the number of times each ad image can be used to 20.

\subsection{Evaluation and Results}
\paragraph{Datasets.} The training set contains 2{,}901 images, of which 2{,}600 have ads. 1{,}600 images with ads were obtained with placeholder replacement, 800 with placeholder replacements with AdChoices logos, 100 with background ads, and 100 with interstitials.

The evaluation set contains a total of 2{,}684 images---2{,}585 with ads and 99 without ads. These are 1{,}595 images with placeholder replacement, 790 images with placeholder replacement with AdChoices logos, 100 images with background ads, and 100 images with interstitials. We also compiled a second evaluation set from 10 domains that were not used for training (this set is different from the one used to evaluate attacks in Section~\ref{sec:attacks}). For each domain, we took a screenshot of the front page and four screenshots of different subpages, resulting in 50 screenshots overall with a total of 75 advertisements. We trained using the default configuration of YOLOv3~\cite{yolo3}, adapted for a unary classification task.


\paragraph{Accuracy and performance.} We tested our model against both evaluation sets. The model achieved the best results after 3{,}600 training iterations. In the first set, our model achieved a mean average precision of $90.88\%$, an average intersect of union of $84.23\%$ and an F1-score of $0.96$. On the second set, our model achieved a mean average precision of $87.28\%$, an average intersect of union of $77.37\%$ and an F1-score of $0.85$. A video demonstrating our model detecting ads on five never seen websites is available at 
{\url{https://github.com/ftramer/ad-versarial/blob/master/videos}}.

We evaluate performance of the model in TensorFlow 1.8.0 with Intel AVX support. On an Intel Core i7-6700 CPU the prediction for a single image took 650ms.

\begin{figure}[t]
	\centering
	{%
		\setlength{\fboxsep}{0pt}%
		\setlength{\fboxrule}{1pt}%
		\fbox{%
			\includegraphics[width=0.7\columnwidth]{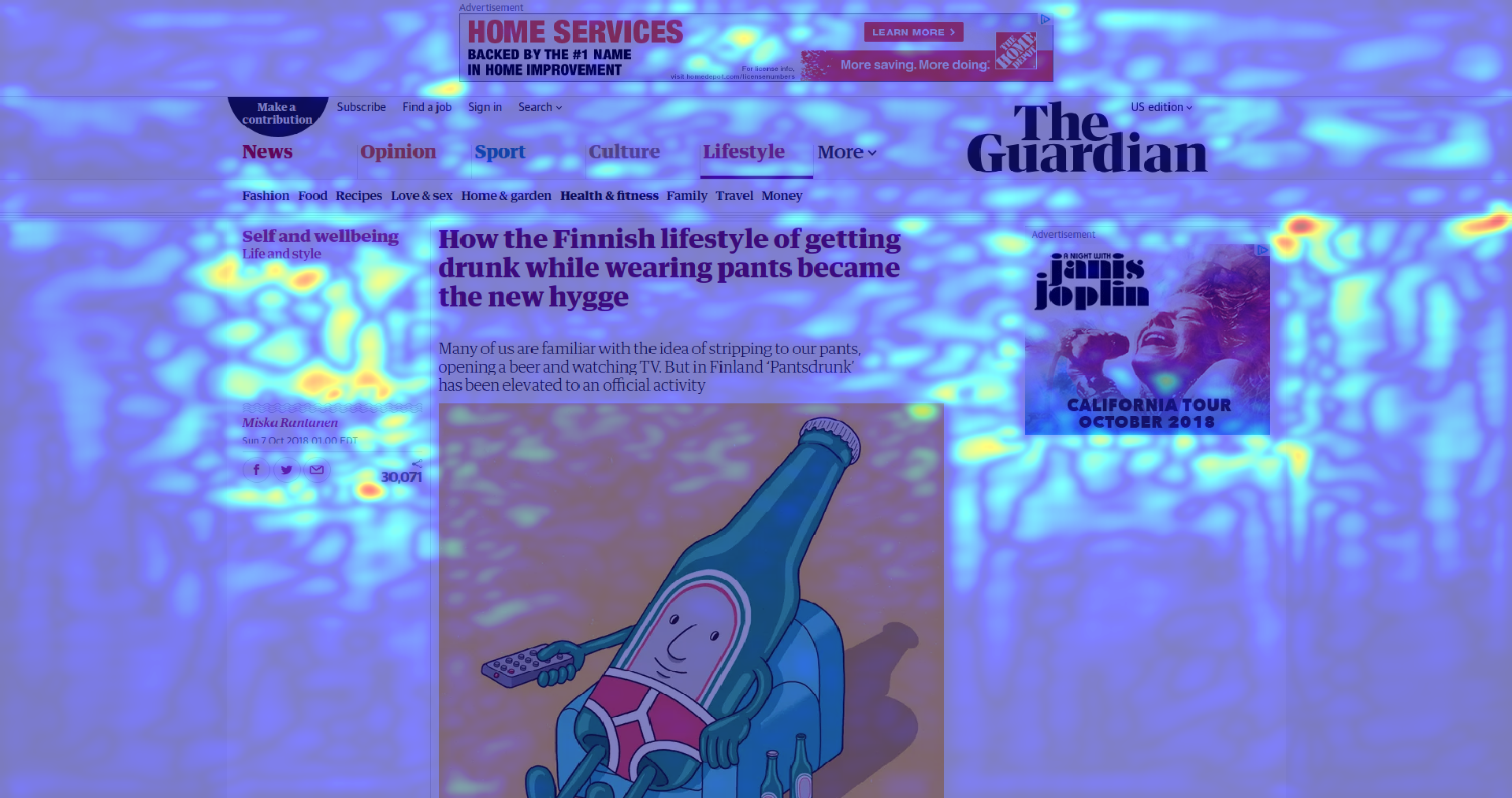}%
		}%
	}%
	\vspace{-1em}	
	\caption{\textbf{Activation Maps of our Ad Detection Model.} The most salient features appear to be the surroundings of ads rather than their visual content.\\[-2em]}
	\label{fig:grad_cam}
\end{figure}

\paragraph{Inspecting our model.} We conduct a preliminary study of the inner-workings of our neural network. By inspecting the model's \emph{activation map} on different inputs (see Figure~\ref{fig:grad_cam}), we find that the model mainly focuses on the layout of ads in a page, rather than their visual content. This shows 
that our ad-blocker detects ads using very different visual signals than humans. This raises an intriguing question about the Sentinel model of Adblock Plus~\cite{sentinel}, which was trained solely on Facebook data, where ads are visually close to the website's native content. Thus, it seems less likely that Sentinel would have learned to detect ads using layout information.

To generate the map in Figure~\ref{fig:grad_cam}, we compute the absolute value of the gradient of the network's output with respect to every input pixel, and apply a smoothing Gaussian kernel over the resulting image. The gradient map is then overlaid on the original input. 

%% file: appendix_extra.tex
\section{Extra Tables and Figures}
\label{apx:extra}

\begin{figure}[!b]
	\begin{minipage}{0.49\columnwidth}
		\begin{subfigure}{\textwidth}
			\centering
			\includegraphics[width=\textwidth]{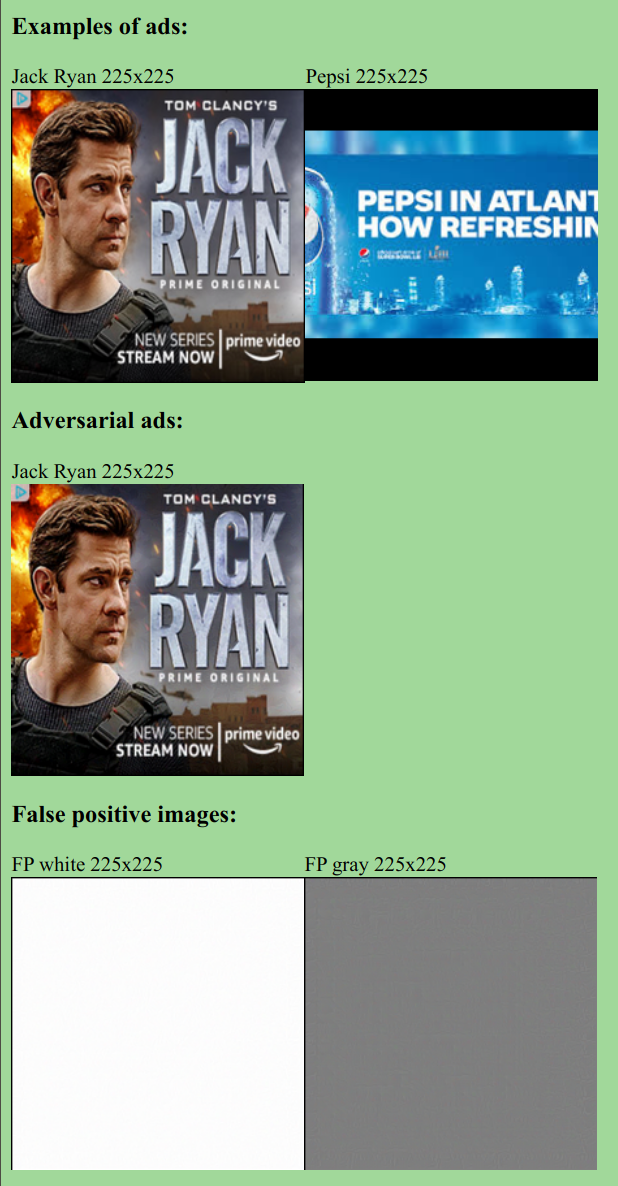}%
			\caption{\textbf{Page displayed in Chromium.}}
		\end{subfigure}
	\end{minipage}
	\hfill
	\begin{minipage}{0.49\columnwidth}
		\begin{subfigure}{\textwidth}
			\centering
			\includegraphics[width=\textwidth]{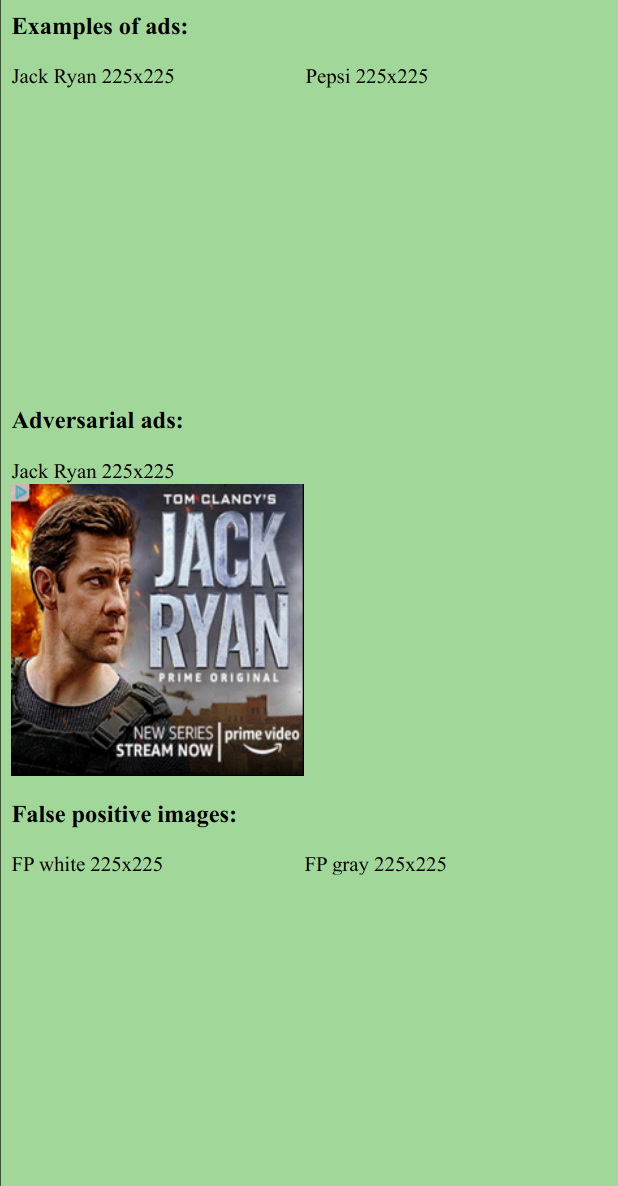}%
			\caption{\textbf{Page displayed in Percival.}}
		\end{subfigure}
	\end{minipage}

	\caption{\textbf{Attack on the Percival Browser from~\cite{percival}.} On the left, a dummy web page is displayed in the standard Chromium browser with two ads (top), an adversarially perturbed ad (middle) and two adversarial opaque boxes (bottom). On the right, the same page is displayed in the Percival browser. The two unperturbed ads on top are correctly blocked, but the adversarial ad evades detection, and the adversarial opaque boxes are mistakenly blocked.}
	\label{fig:percival-attack}
\end{figure}

\begin{table}[!b]
	\caption{\textbf{Evaluation Data for Adversarial Examples.} We collect images, frames and screenshots from the Alexa top ten news websites that use the AdChoices standard (we exclude \url{news.google.com} and \url{shutterstock.com} which contain no ads on their front-page).
		For each page, we extract all images below 50 KB, all iframes, and take two screenshots (the front page and an article) of the user's viewport, and report the number of visible ads in these.\\[-2em]}
	\centering
	\setlength{\tabcolsep}{2.5pt}
	\footnotesize
	{\fontsize{7.5}{0}%
		\def\arraystretch{0.8}
		\begin{threeparttable}
			\begin{tabular}{@{}l r r r r r r@{}}
				&
				\multicolumn{2}{c}{\textbf{Images}}
				&
				\multicolumn{3}{c}{\textbf{Iframes}}
				&
				\\[-0.7em]
				&&&&&&\textbf{Visible}\\[-0.7em]
				\cmidrule(lr){2-3}\cmidrule(lr){4-6}
				\textbf{Website} & 
				\textbf{Total} &
				\textbf{AdChoices} &
				\textbf{Total} &
				\textbf{Ads} & 
				\textbf{AdChoices} & 
				\textbf{Ads}
				\\
				\toprule 
				\url{reddit.com} 		&70 &2 & 2 &2 &2 &2\\ 
				\url{cnn.com} 			&36 &7 & 7 &5 &2 &3\\
				\url{nytimes.com} 		&89 &4 & 3 &3 &3 &2\\
				\url{theguardian.com} 	&75 &4 & 8 &3 &3 &3\\
				\url{indiatimes.com} 	&125&4 & 5 &5 &4 &3\\
				\url{weather.com} 		&144&5 & 11&7 &3 &3\\
				\url{news.yahoo.com} 	&100&5 & 3 &3 &2 &3\\
				\url{washingtonpost.com}&40 &1 & 5 &2 &1 &3\\
				\url{foxnews.com} 		&96 &5 & 6 &5 &4 &4\\
				\url{huffingtonpost.com}&90 &4 & 9 &4 &5\tnote{$\dagger$} &4\\
				\midrule
				\textbf{Total} 			&865&41& 59&39&29&30
			\end{tabular}
			\begin{tablenotes}
				\item[$\dagger$]One AdChoices logo appears in two rendered \texttt{iframes} laid on top of each other.
			\end{tablenotes}
		\end{threeparttable}
	}
	\setlength{\tabcolsep}{6pt}
	\label{tab:websites}
	\vspace{2em}
\end{table}

\begin{figure}[!b]
	\begin{minipage}{\columnwidth}
		\begin{subfigure}{\textwidth}
			\centering
			{%
				\setlength{\fboxsep}{0pt}%
				\setlength{\fboxrule}{1pt}%
				\fbox{%
					\includegraphics[width=\textwidth-2pt]{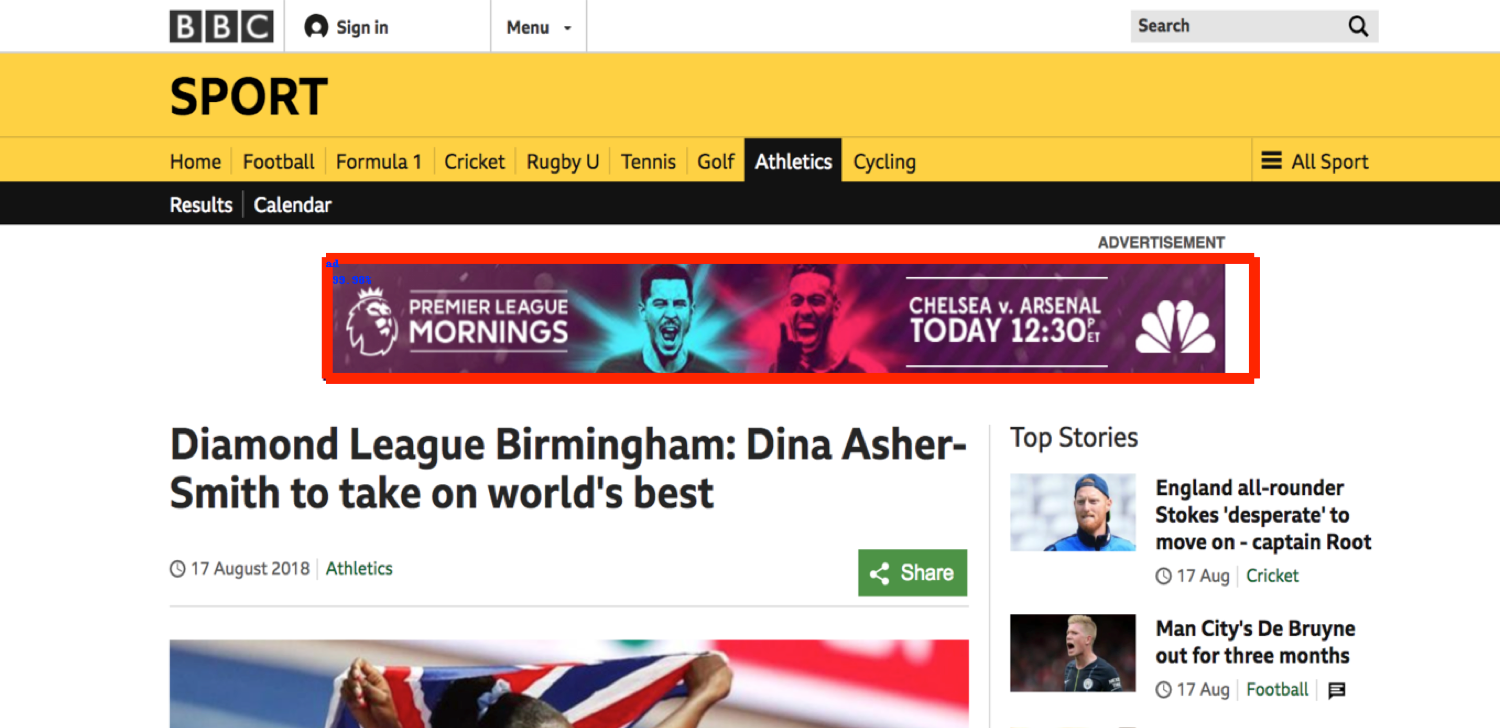}%
				}%
			}%
			\caption{\textbf{Original Page:} The ad banner is correctly detected.}
			\vspace{0.5em}
		\end{subfigure}
	\end{minipage}
	\\
	\begin{minipage}{\columnwidth}
		\begin{subfigure}{\textwidth}
			\centering
			{%
				\setlength{\fboxsep}{0pt}%
				\setlength{\fboxrule}{1pt}%
				\fbox{%
					\includegraphics[width=\textwidth-2pt]{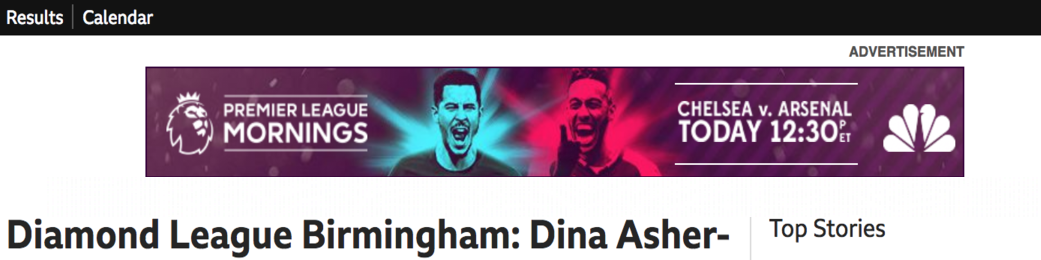}%
				}%
			}%
			\caption{\textbf{Attack C3-C4:} The publisher perturbs the white background beneath the ad to evade ad-blocking (C4). Alternatively, an ad network adds a universal mask on the ad (C3, not displayed here for brevity). In both cases, the perturbation is invisible to the user.}
			\vspace{0.5em}
		\end{subfigure}%
		\\
		\begin{subfigure}{\textwidth}
			\centering
			{%
				\setlength{\fboxsep}{0pt}%
				\setlength{\fboxrule}{1pt}%
				\fbox{%
					\includegraphics[width=\textwidth-2pt]{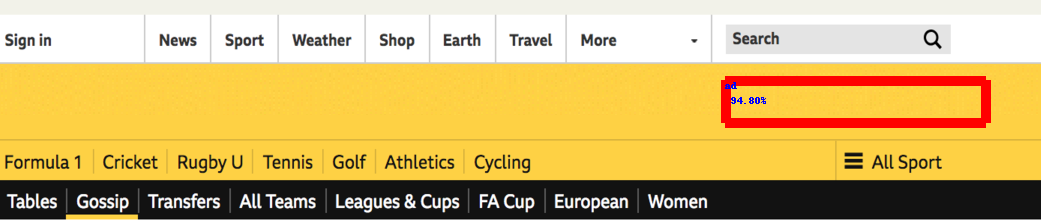}%
				}%
			}%
			\caption{\textbf{Attack C1:} The publisher adds a honeypot element to the page header (top-right) to detect an ad-blocker.\\[-2em]}
		\end{subfigure}
	\end{minipage}
\vspace{-0.3em}
	\caption{\textbf{Universal Adversarial Examples for Page-Based Ad-Blockers on \url{BBC.com}.}  Examples of evasion attacks C3-C4 and detection attack C1 (see Section~\ref{ssec:attacks_classification}).}
	\label{fig:bbc-apx}
\end{figure}